\pdfoutput=1
\documentclass[traditabstract]{aa}

\usepackage{graphicx}
\usepackage{txfonts}
\usepackage{natbib} 
\bibpunct{(}{)}{;}{a}{}{,} % to follow the A&A style

\begin{document}

\title{Approximate integrals of motion}

\author{Olivier Bienaym\'e\inst{1} and Gregor Traven\inst{2}}

\institute{Observatoire Astronomique de Strasbourg, UMR 7550 CNRS / Universit\'e de Strasbourg 
	\and
	University of Ljubljana, Faculty of Mathematics and Physics}

%\offprints{ \email{olivier.bienayme@unistra.fr}}

\date{Received / Accepted }

\abstract{We determine  approximate numerical integrals of motion of 2D symmetric  Hamiltonian systems.  We detail for a few gravitational potentials the conditions  under which  quasi-integrals are obtained as  polynomial series. We show that  each of these potentials  has a wide range of  regular orbits  that are accurately modelled with a unique approximate integral of motion.}{}{}{}{}


\maketitle

\section{Introduction}

Gravitational potentials with explicitly known  second integrals of motion are  rare, the most frequently used  in astronomy are  axisymmetric potentials and also \cite{sta93} potentials\footnote{St\"ackel  potentials, as they are called  in the  astronomical literature, are   usualy named Darboux potentials in the mathematics literature. A first systematic attempt to study  quadratic invariants is given in \cite{dar01}.  Contributions on separability, orthogonal coordinate systems, and quadratic invariants may be found in \cite{ank83}. }
that cover  a large class of potentials, but are not always sufficiently realistic. Such explicit forms are useful, for instance, to understand more clearly the underlying physics of dynamical systems, to build stationary distribution functions, etc. \cite{hie87}  gives such a list of other potentials with known second integrals, but potentials with an analytic second integral of motion are  the exception as shown by Poincar\'e, and ergodic motions should be the rule. However, according to the KAM theorem,  at low energies many orbits are regular and remain confined on tori, while for irregular orbits the time diffusion may be exponentially long \citep{nek77,mor95a, mor95b}. For these orbits, as shown  numerically by \cite{oll62} for a realistic galactic potential, we expect the existence of approximate first integrals.

The need for  generality leads to numerous  efforts to obtain tractable integrals of  motion for more   realistic potentials. Among the developed methods, we  can mention the dynamical spectra and more specifically  the frequency analysis  \citep{bin82,las93} with a recent 3D application to numerical simulations of galaxies \citep{val12}. Frequencies associated to each orbit are constants of motion and allow  accurate classifications of orbits. Another example, closer to the work developed here, is the tori reconstruction of \cite{mcg90}, who modelled  the tori on which orbits circulate in the phase space.  Their analysis is based on  the angle-action variables, and the time dependency remains explicit. This method is  very attractive since   the dynamics of the system becomes very simple. They distort analytic tori of a  toy Hamiltonian into the tori of the Hamiltonian of interest. This allows the analysis of  any non-integrable potential as a perturbation of a nearly integrable one  \citep{bin93}. Extensions for the separate modelling of orbit families are described by \cite{kaa94a}, \cite{kaa94b,kaa95}. More direct modellings of orbits are also proposed by \cite{pre82} using the ratio of trigonometric functions, or by \cite{rob93}  using Pad\'e approximations (rational functions). 
Close to our work, \cite{war91}  determined approximate integrals by fitting orbits. He developed a fit using  an approximate discrete Fourier transform of positions along orbits and  recovered the coefficients of an  integral of motion.  This method works for non-resonant or high-order resonant orbits but fails for  low-order resonant orbits.
\cite{baz91}  determined an approximate integral for (2D) sympletic maps,  minimising  the variation of the approximate integral along the  orbit under study.

Formal integrals of motion  can be obtained by directly solving   the Boltzmann equation assuming that the integral may be written as a polynomial  \citep[see][]{con60}. The convergence of these   series is not guaranted, but  they may be asymptotic  (i.e. semi convergent series), allowing one to  approximate integrals. \cite{bir27}  and \cite{gus66}, using normal forms, proposed a process to  build such approximate formal integrals of motions. A more complete description of all these subjects can be found in  \cite{con02}. Introductory lectures  on these questions of galactic dynamics and celestial mechanics may be found in \cite{hen83} or \cite{eft07}.
  
  The motivation of the work presented in this paper is to obtain tractable and approximate integrals of motion for a few simple potentials that are representative of galactic potentials.  We restricted our study to 2D motions. 
  Future applications might be the building of dynamically self-consistent galactic models for stationary potentials, and modelling the kinematics of its stellar populations with  distribution functions. These goals are usually achieved with N-body numerical simulations, with the   \cite{sch79}, or with the \cite{sye96}   methods.
 
Our method consists of solving the stationary Boltzmann equation for a second integral of motion.  Instead of using a formal integration, we numerically determine the coefficients of a series by a least-squares minimisation of the integral  variance along orbits. This allows us to obtain a single expression for different orbit families. The method is described in Section\,2 and detailed explanations and discussion for a simple 2D potential are given in Section\,3. Applications to other potentials are briefly described in Sections\,4-5, and the conclusion is presented in Section\,6.

\section{Method and algorithm}

According to a definition proposed by \cite{hie87}, the concept of integrability is to be able to make some quantitative general statements of the dynamical system under study using a quantity whose value is conserved during the time evolution of a system.

In this paper we  approach this problem numerically and consider  two degrees of freedom  time independent Hamiltonian systems of the form
$$H=\frac{1}{2} (u^2+v^2)+\Phi(x,y),$$

with $x,y$  the space coordinates and $u,v$ their conjugate momenta.
The energy is naturally a first invariant, and  we search numerically for a second invariant 
$I(\vec{x},\vec{v})=\tilde{I}(x,y,u,v)$ independent of the energy.

We recall that this problem is closely related to the collisionless Bolztmann equation and that any distribution function independent of time of the form $f(\vec{x},\vec{v})=g(I(\vec{x},\vec{v}),E(\vec{x},\vec{v}))$ will be the solution of  the stationary 2D Boltzmann equation,

$$ u \frac{ \partial f}{\partial x} + v \frac{  \partial f}{ \partial y}  -\frac{\partial\Phi}{ \partial x} \frac{\partial  f}{\partial u}   
 -\frac{\partial\Phi}{ \partial y} \frac{\partial  f}{\partial v}  =0, $$

since this equation is also the Poisson bracket $\{f,H\}$=0 that is a property satisfied by any invariant of the Hamiltonian systems.

From  Hamilton's equations of motion, we can determine orbits and note that if a second invariant  $I(\vec{x},\vec{v})$ independent of the energy exists,  it  remains constant, by definition,  along any   regular orbit. Thus we  assume the existence of  such a second invariant, and we  also assume that it may be written as a polynomial finite series of the coordinates and momenta $(x,y,u,v)$:
\begin{equation}
I=I(\vec{x},\vec{v})=\sum_{k,l,m,n}  c_{k,l,m,n} \,x^k y^l u^m v^n\,,
\end{equation}
To determine the coefficients of the series, we select a set of regular orbits for a given potential, for each orbit a set of positions ($x,y,u,v$) along  that orbit, and for each position $I$ is evaluated.  Coefficients are  computed by a least-squares minimisation that    minimises  the variation of $I$ around its mean value, (i.e. minimising the standard error of $I$) along each orbit. This standard error would be zero if an exact integral $I$ existed for the considered potential  and, according to our construction here, if it were also a polynomial. 
Before realising the minimisation, a number of considerations  allows us to cancel redundant and useless coefficients within Eq.\,1. First, to avoid  obtaining the trivial solution where all  coefficients are null,  we  fix a coefficient to the value one: here, we will fix the coefficient of one these terms: either $x^2$, $v^2$, or $x^2v^2$. Many degrees of freedom remain to determine the coefficients  within the series of $I$, and we may want that the process to build $I$ leads to a unique solution. The unicity of  adjusted solutions is necessary and useful to allow the comparison of solutions obtained using different orbits, different  integration time for orbits, different number of coefficients, etc.  We also want that the integral of motion $I$ is independent of the energy. For that purpose,  we  modify $I$ by removing the term $u^{2m}$ from the series by subtracting from $I$ a quantity proportional to $E^m$ (thus the modified $I$ remains an integral). Starting with $m=1$ and increasing $m$, we iteratively modify $I$, removing   all   terms $u^m$, and now   $I(\vec{x},\vec{v})$  and $E(\vec{x},\vec{v})$ are independent. All this  is to explain that we do not remove useful terms, because the independence is satisfied   if the rank of the Jacobian $\partial(E,I)/\partial(x_i,p_j)$ is 2  (instead of removing the $u^m$ terms, we could have cancelled  the coefficients of the $v^{2n}$ terms, or as efficiently   those of the $x^k$ or $y^l$ terms.)

Some other coefficients are  redundant. If $I$ is an invariant, so is $g(I)$, where $g$ is any analytic function. To ensure that our minimisation process produces a unique solution,  we reiterate and  now subtract from $I$ a quantity proportional to $I^k$ ($k >1$)   to  remove the  term $x^k$. Iterating this 
operation\footnote{Here, we do not answer the question of the convergence of these iterative operations. Our success in obtaining approximate integrals in the next paragraphs indicates that the iterations can build at least asymptotic series.}, 
all terms of the form $x^k$,  except the first one, are removed from the series. If the potential is even for the $x$ coordinate, the first remaining term  is in general the $x^2$ term.
We find that   selections of the coefficients and energy independence are critical for obtaining a numerically accurate second integral of motion.

Finally, a significant number of other coefficients are removed for symmetry reasons. Each of the $x$ and $y$ axial symmetry of the potentials considered in this paper allows us to cancel half of the coefficients. Moreover, it is also known \citep{hie87} that if $I$ is an integral of motion, it can be split according to the momenta parity, and the odd and even part  $I_+$ and  $I_-$ are also integrals. If the potential is not superintegrable (and thus has at most two integrals), $I_+$ or $I_-$ must be equal to zero (otherwise they are dependent), and then the momenta parity of $I$ is clearly defined (in practice if the potential contains box orbits, the momenta parity of $I$ is even.)

Practically, we rewrite the polynomial $I$, where the $a_i$ are $N$ distinct monomials and the  $c_i$ are $N$ adjusted coefficients, as

$$I=a_0+ \Sigma_{i=1}^N \,c_i\, a_i\, .$$

We set, for instance, $a_0=x^2$ and its coefficient $c_0$ is  1.

From $M$ positions  along a given orbit,  we set $a_{i,m}=a_i(x_m,y_m,u_m,v_m)$,    with $m=1$ to $M$.

We define the mean value of $I$ over the $M$ positions as 

$\bar I=\frac{1}{M} \Sigma_{m=1}^M I_m$

and its standard deviation as

$\sigma^2=\frac{1}{M} \Sigma_{m=1}^M (I_m-\bar I)^2$.

Minimising the standard deviation reduces to  $N$ linear equations equivalent to the matrix equation 

$$D . C + B=0$$

with $$C_i=c_i\, ,$$

$$ B_i=\Sigma_{m=1}^M \alpha_{0,m}. \alpha_{i,m}\,,$$

$$D_{i,j} =  \Sigma_{m=1}^M \alpha_{i,m} . \alpha_{j,m} $$

and 

$$ \alpha_{i,m}=[   \,a_{i,m}  - \frac{1}{M}\Sigma_{m''=1}^M \,a_{i,m''} ]\,.$$

In the case of simultaneous fitting of $P$ different orbits with $M$ positions along each orbit, the linear equations remain similar. If the index $m$ and $p$ refer to the position $m$ on orbit $p$, we have

$$ B_i= \Sigma_{p=1}^P \,B_{i,p} = \Sigma_{p=1}^P
 \alpha_{0,m,p}. \alpha_{i,m,p}\, ,$$

$$D_{i,j} =   \Sigma_{p=1}^P D_{i,j,p} =
 \Sigma_{p=1}^P  \Sigma_{m=1}^M \alpha_{j,m,p} . \alpha_{i,m,p} $$

and

$$ \alpha_{i,m,p}=[   \,a_{i,m,p}  - \frac{1}{M}\Sigma_{m''=1}^M \,a_{i,m'',p} ]\,.$$

When $M \times  P$ is larger than $N$, the problem is overdetermined,  and we use the {\it dgesv} routine (LU decomposition) from the LAPACK software (www.netlib.org/lapack/)   to solve the system of linear equations. Other LAPACK routines have been tried and give  similar accuracies.

The various    fixed or removed coefficients depend on the shape of the potential.
In Sections 3, 4.1, and 4.2 we set the following coefficients:

$c_{2,0,0,0}=1$

$c_{2k,0,0,0}=1$ with $k>1$

$c_{0,2l,0,0}=0$ with $l \ge 1$

$c_{k,l,m,n}=0$ if $l+m$ or $k+n$ or $m+n$ is odd.

In Section 4.3  the following coefficients  are fixed:

$c_{0,0,0,2}=1$

$c_{0,0,0,2n}=1$ with $n>1$

$c_{0,0,2m,0}=0$ with $m \ge 1$

$c_{k,l,m,n}=0$ if $l+m$ or $k+n$ or $m+n$ is odd.

In Section 5  the following coefficients  are fixed:

$c_{2,0,0,0}=1$

$c_{0,0,0,n2}=0$ with $n > 1$

$c_{2k,0,0,0}=1$ with $k \ge 1$

$c_{k,l,m,n}=0$ if  $k+n$ is odd or $m+n$ is even.\\

In summary, the method presented here to solve the Boltzmann equation and to obtain a first integral  $I(\vec{x},\vec{v})$  consists of  computing  some regular orbits,  and of adjusting $I(\vec{x},\vec{v})$   to these peculiar solutions. The constraint on the coefficients of the polynomial series modelling  $I(\vec{x},\vec{v})$   is that $I$ remains constant along each selected orbit. The most important details are given within the next section using an 'exponential' potential. Other results are summarised for a few potentials in the following sections.

\begin{figure}
\resizebox{\hsize}{!}{\rotatebox{270}{\includegraphics{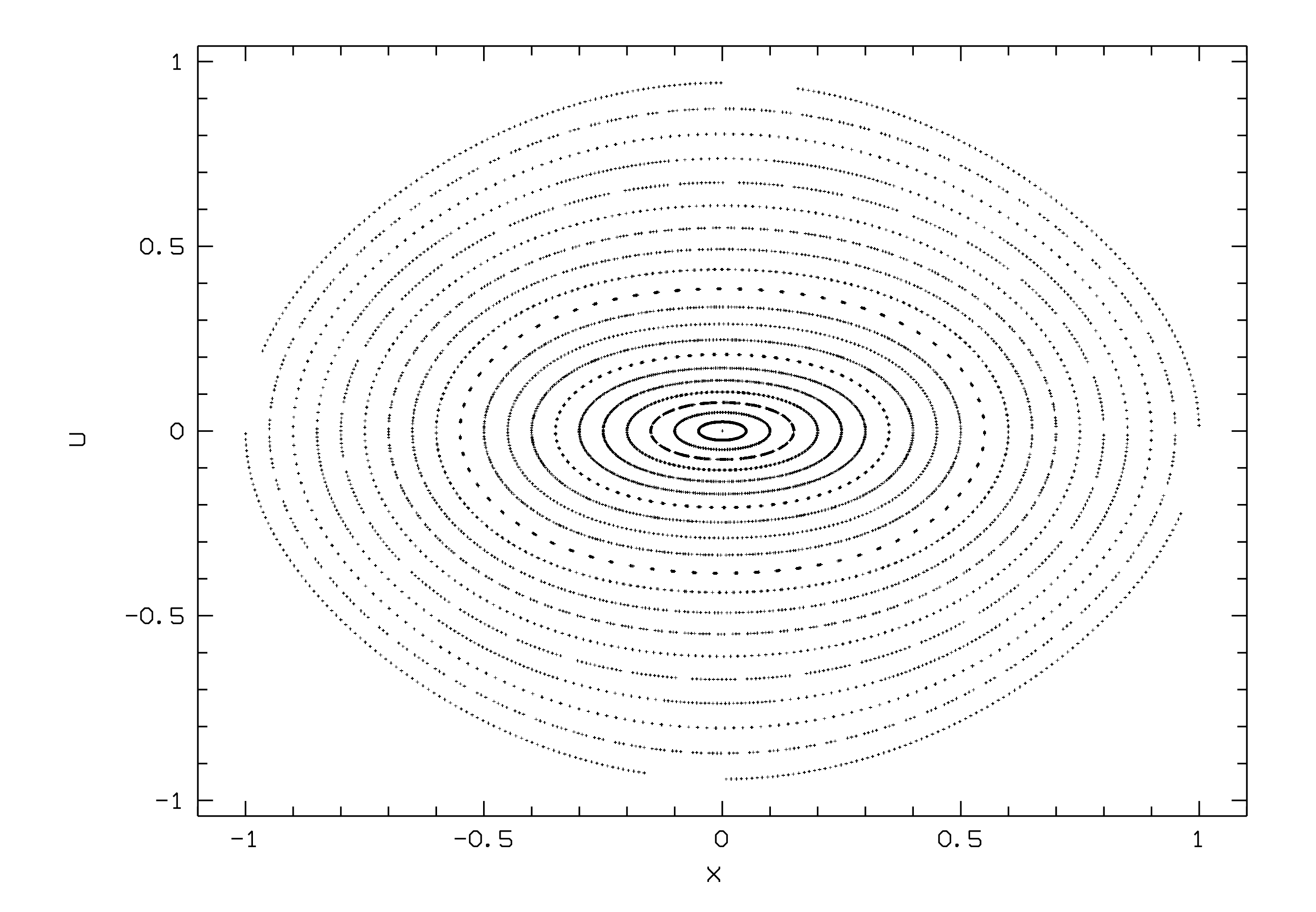}}}
\resizebox{\hsize}{!}{\rotatebox{270}{\includegraphics{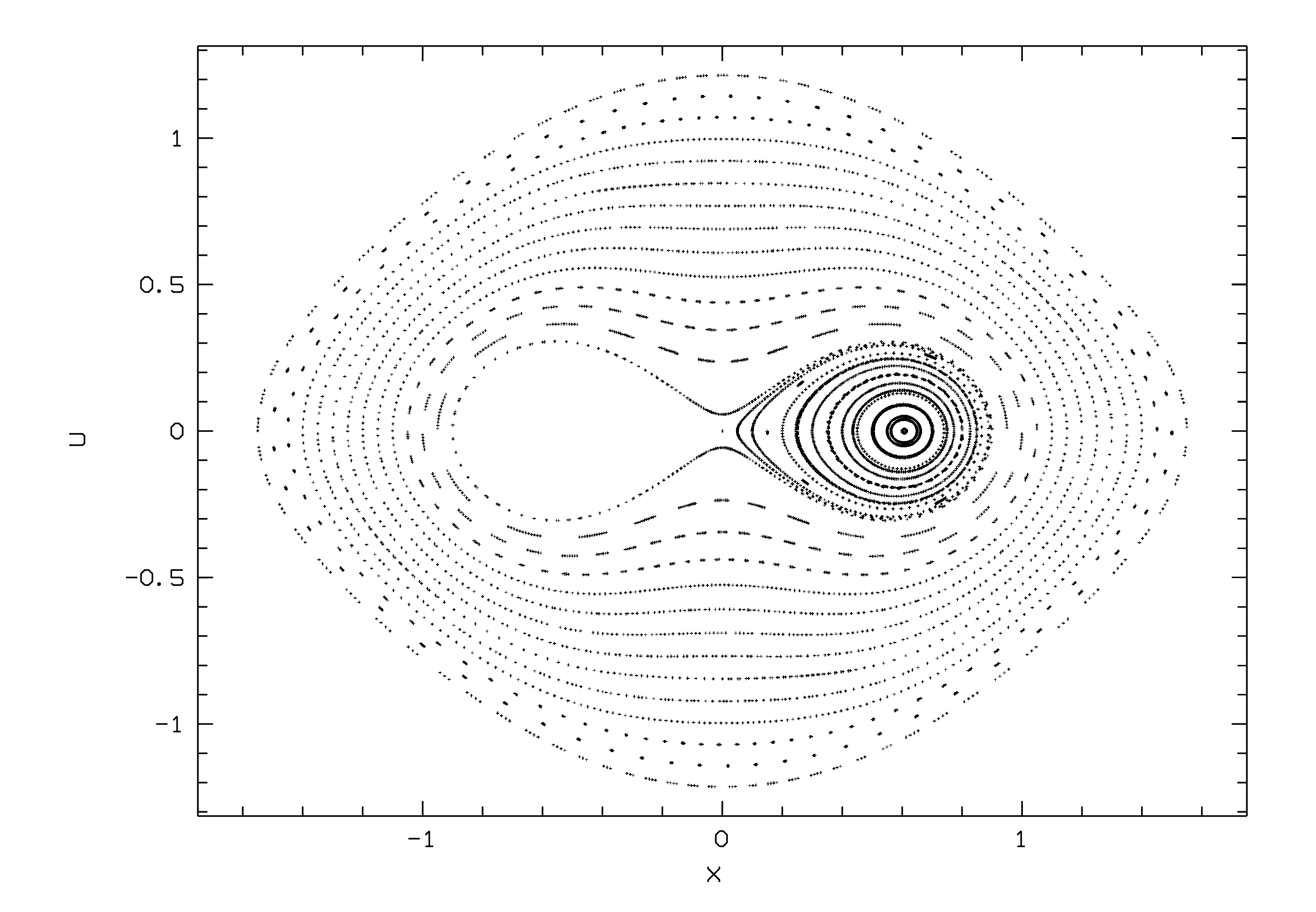}}}
\caption{Poincar\'e surfaces of section for the "exponential" potential  with $q=0.8$ and energies $E=-0.5$ (top) and  -0.2 (down).}
  \label{fig1}
\end{figure}

\begin{figure}
\resizebox{\hsize}{!}{\rotatebox{270}{\includegraphics{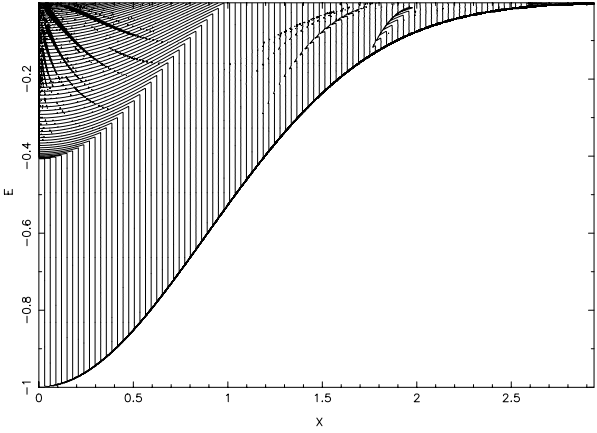}}}
  \caption{Isolevels of $x$-axis orbit apocentres versus $x_{\rm init}$ and $E$. At low energies box orbits dominate the diagram. Tube orbits (type 1:1) become significantly present at $E  >   -0.4$. At $E=-0.05$,  they cover the range  $x_{\rm init}$ from 0 to 1.4. Higher order resonances draw visible "plumes". }
 \label{fig2}
\end{figure}

\section{Exponential potential}

We  apply  the method described in Section\,2  to obtain an approximate integral of motion to the case of the following barred potential:
$$\Phi(x,y)=-\exp(-\,q^2\, x^2 -y^2).$$

This   potential  is chosen for its simplicity: bounded orbits have energies in a limited range within [-1,0] and most orbits are regular. The density associated to the potential is not positive everywhere but  the Taylor series of this potential has an infinite radius of convergence that should allow us, in a future work, an easier  comparison of our results to formal integrals obtained by other means \citep{gio78}.

Most of the bounded orbits ($E<0$) are regular. They are  box and tube orbits (libration and circulation), and most of the tube orbits are of type (1:1). This can be seen in Poincar\'e surfaces of section (Figs.~\ref{fig1}) or  with the 2D isolevel plot of  x-apocentre (at $u=y=0$)  versus the energy and  the initial position $x_{\rm init}$ (at $u=y=0$) (Fig.~\ref{fig2}). In this last figure, high-order resonances appear at the highest energies   and draw inclined  'plumes'. 

To build the polynomial quasi integral of motion, we  select  orbits   that cover a wide range of energies and  a wide range of second constants  of motion, taken as the orbit initial position $x_{\rm init}$ ($u=y=0$) (in this  example, we  consider only orbits that cross the x-axis perpendicularly,   which excludes  very few orbits  that appear mainly at  high energies.) 

We also select  the polynomial terms used to determine $I$.  Due to  the $x$- and $y$-axis symmetries of the potential,  the coefficients of terms  in Eq.~1 with  $l+m$ or $k+n$ odd  are set to zero. The momenta parity of $I$ is even ($m+n$ even). We set the coefficient of $x^2$ to one, remove all $x^{2k}$ ($k>1$) and all $v^{2l}$ terms. The final series thus obtained will be independent of $E$ and a unique solution for the coefficients is expected. 

The remaining coefficients of the series in Eq.~1  are obtained by a least-squares minimisation, and the quality of the fit is quantitatively defined by the constancy of $I$ along each orbit. The result of the fit critically depends on the existence or non-existence for the examined potential  of an approximate integral that remains nearly constant  over long periods of time along orbits.  But the quality,  also critically, depends as any polynomial fit, on a convenient coverage of the fitted space to adjust structures, and on a sufficiently large number of coefficients to model the numerous and small structures. Finally, it also depends on  a sufficiently large number of fitted data to avoid or at least to minimise  erratic oscillations between fitted data. For this,  we impose   that the number of fitted positions is sufficiently large by applying  the recommended conditions that $2\,N_{\rm pos}^{1/2} \gg N_{\rm coeff} $ \citep{dal74}.  We must also ensure a sufficient coverage:  1000 positions on each orbit corresponds to a mean distance of 11 $\deg$ for the angle positions, assuming   the couple of angles and actions were known.
 
Orbits are computed with a fourth-order Runge-Kutta integrator, the time step is fixed and the energy is conserved at 10$^{-8}$. The minimisation is written as a linear least-squares fit, inversion is performed  using the LAPACK softwares and presents no peculiar difficulties, because the matrix size remains small. \\

% figure3

\begin{figure}
\resizebox{\hsize}{!}{\rotatebox{270}{\includegraphics{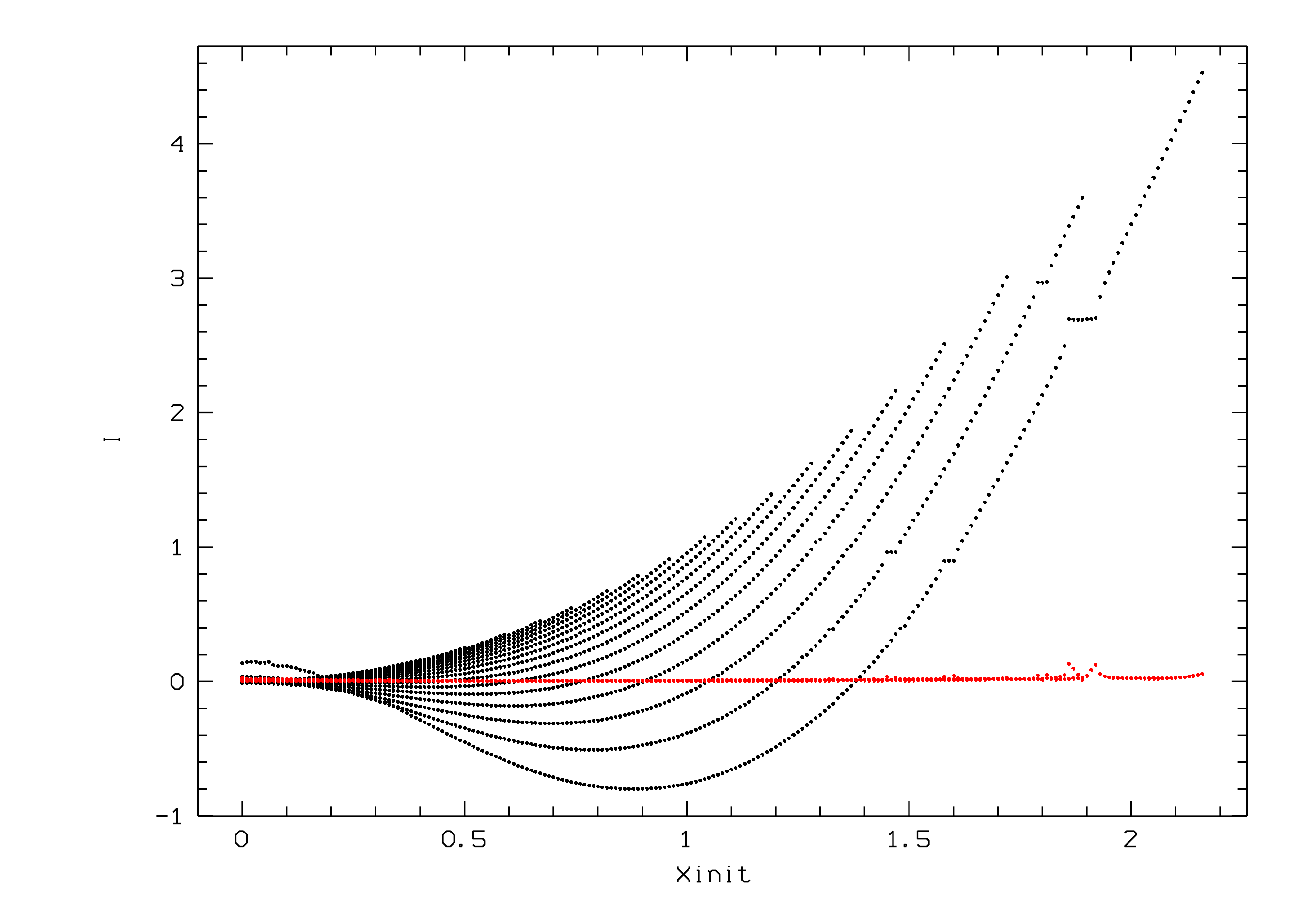}}}
  \caption{ Second integral of motion $I$ (polynomial of order 12) versus $x_{\rm init}$ for 2085 orbits covering 19  energies.   Orbits with the same energy draw 19 dotted lines. Box orbits have $I>0$, tube orbits have $I<0$. The red continuous line is the dispersion of $I$, it increases at resonances.}
  \label{fig3}
\end{figure}

% figure 4

\begin{figure}
\resizebox{\hsize}{!}{\rotatebox{270}{\includegraphics{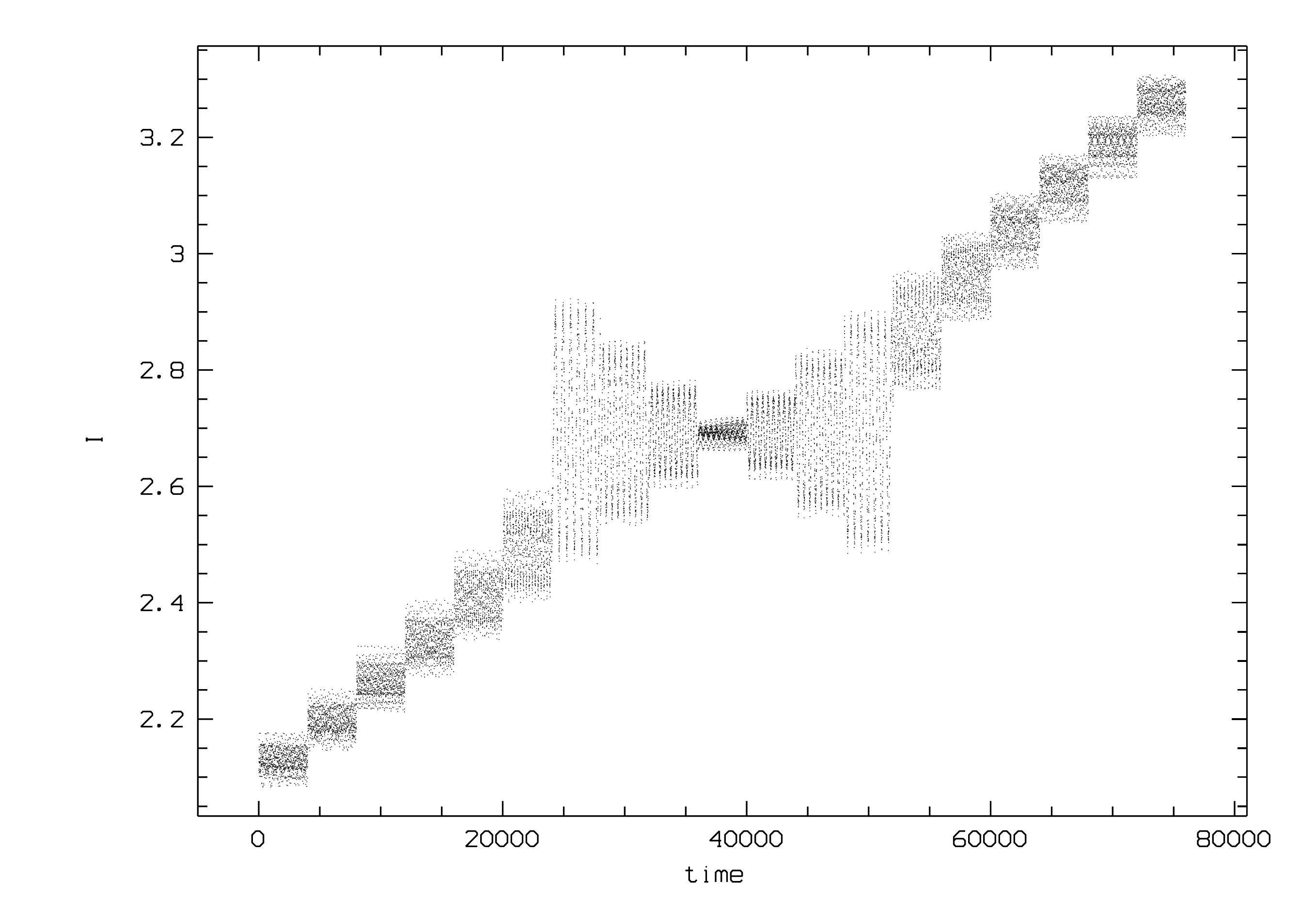}}}
 \caption{ For 19  orbits with $E=-0.05$ close to the resonance 2:3, $I(t)$ is plotted every $\Delta t=4$ over the time interval $T=4000$. The dispersion of $I(t)$ increases for orbits close to  the resonance. The time sequence is shifted for each orbit for visibility.}
  \label{fig4}
\end{figure}

\subsection{Results}

The first  example is  obtained with the  following  conditions:
we select  2085 orbits covering energies $E$ = $-0.95, -0.9, ..., -0.05$  and at each energy with initial conditions  ($x_{\rm init}$, $y_{\rm init}$=0, $u_{\rm init}$=0, $v_{\rm init}(E,x_{\rm init})$)  $x_{\rm init}$  covering the interval  0 to $x_{\rm max}(E)$ by step $\Delta x_{\rm init}=0.01$. For each orbit, 1000 positions  are  taken (time step $\Delta T=4$) from $T$ = 0 to $4000$. After  removing  cancelled terms, we use the first 200  coefficients of Eq.~1,  a 12$^{th}$ order polynomial, and  after fitting, following the procedure described in the previous section, we obtain $I_{12}$  a quasi invariant  integral of motion.

  We find that the mean value, $\bar{I}_{12}$,     varies from orbit to orbit between 
  $\sim -1$ and $\sim+4.6$ (Fig.~\ref{fig3}) and  that the mode (histogram maximum) of the dispersions $\sigma_{I_{12}}$ along each orbit is   0.002,  and $\sigma_I$ remains within   $[5.10^{-4} -\, 3.10^{-3}]$ for the low-energy orbits ($E<-0.5$). At higher energies, 30 orbits  close to resonances have   $\sigma_I$ within $[0.03 - 0.15]$, the remaining 1360 orbits have   $\sigma_I$ smaller than 0.03 and a mode equal to $0.002$.

We find that $I_{12}$ remains  nearly invariant  over long periods  by computing for each of the 2085 orbits $\bar{I}_{12}$ over the time interval $\Delta T=4.10^{6}$, corresponding approximately to 50000 rotations, a much longer time than the   interval $\Delta T = 4. 10^3$  used for the fit (50 rotations). The variations of $I_{12}$ (i.e. its residuals) are   identical to a few percent to residuals obtained within the fitted time interval and, thus, we conclude  that $I_{12}$ can be  extrapolated in time far outside the fitted domain (but of course not outside the fitted domain in coordinates and momenta).

We extend this check to examine  the variation of $I_{12}$ for  orbits not used for the fit. Therefore, we explore in detail the  energy-$x_{init}$ fitted domain with a finer grid in energy ($\Delta E=0.01$ and $\Delta x_{\rm init}=0.0025$)   using $\sim$32000 orbits. For these orbits the dispersion remains  very close to the dispersion of the nearby orbits used for the fit. This  confirms that we use a sufficiently large number of orbits and positions and that the interpolation between fitted orbits is correctly achieved.
 
 Figure~\ref{fig3} presents,  for each of the 2085 orbits, the mean value of $I_{12}$ along the orbit versus its initial positions  $x_{\rm init}$.   For each orbit,   $x_{\rm init}$ is either its $x$-pericentre or its $x$-apocentre along the $x$-axis (thus with $y$=$u$=0). Each dotted line represents a sequence of orbits with an identical value of the energy. 

The two main families of orbits are  box orbits and  tube orbits (with type 1:1),  corresponding   to positive and negative values of $I_{12}$.  The family of box orbits is dominant   at low energies, $E \la -0.4$. At all energies, the 1:1 periodic tube orbit  corresponds to the minimum of $I_{12}$. 
For most of the orbits, there is  a one-to-one relation between each orbit and a  couple of values ($E,I$). This useful property is  lost, however, when the fit is improved to adjust higher order resonances,  as we   show below.

% figures 5

\begin{figure*}
\begin{center}
\resizebox{8.5cm}{!}{\rotatebox{270}{\includegraphics{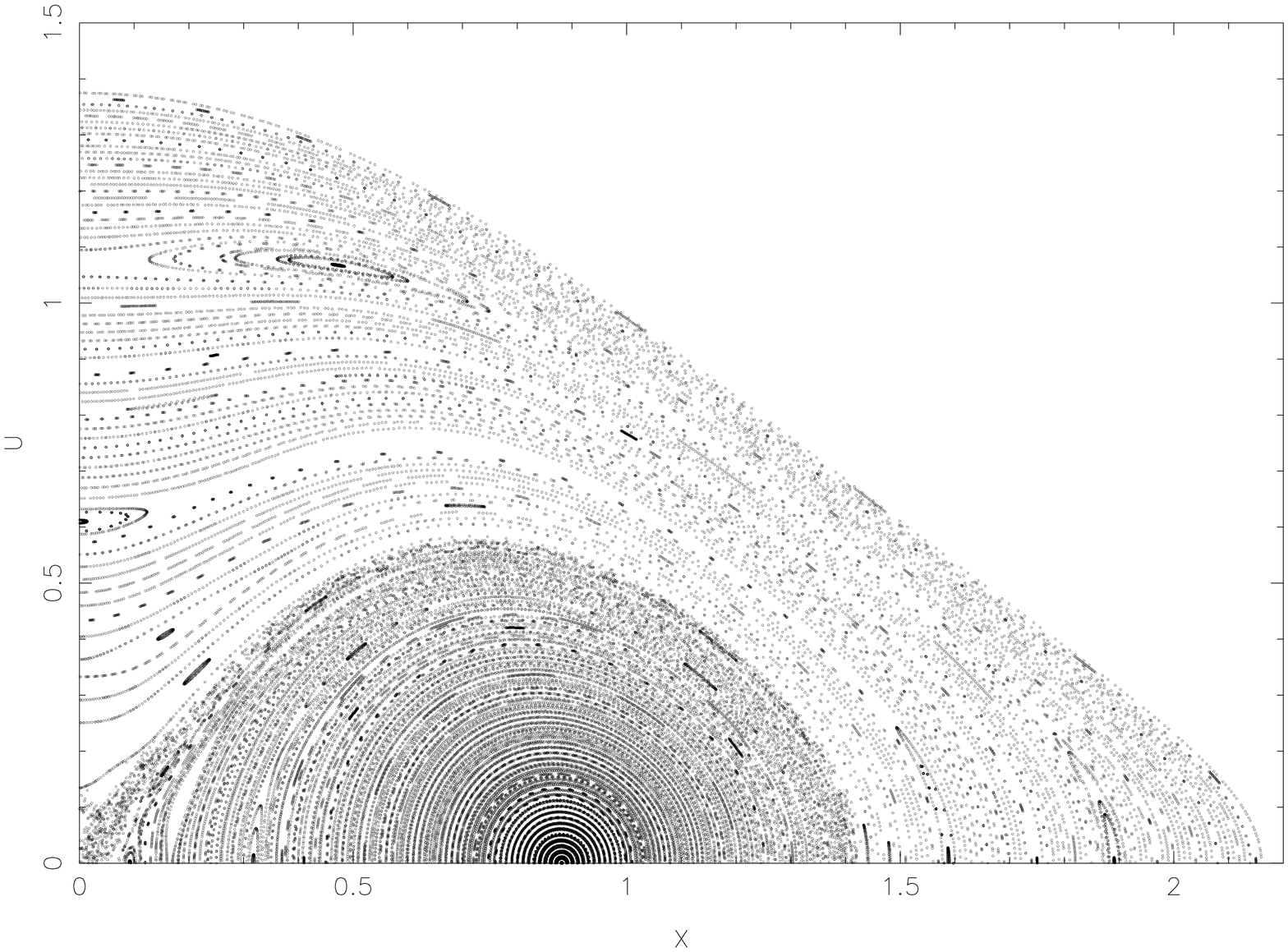}}}
\resizebox{8.5cm}{!}{\rotatebox{270}{\includegraphics{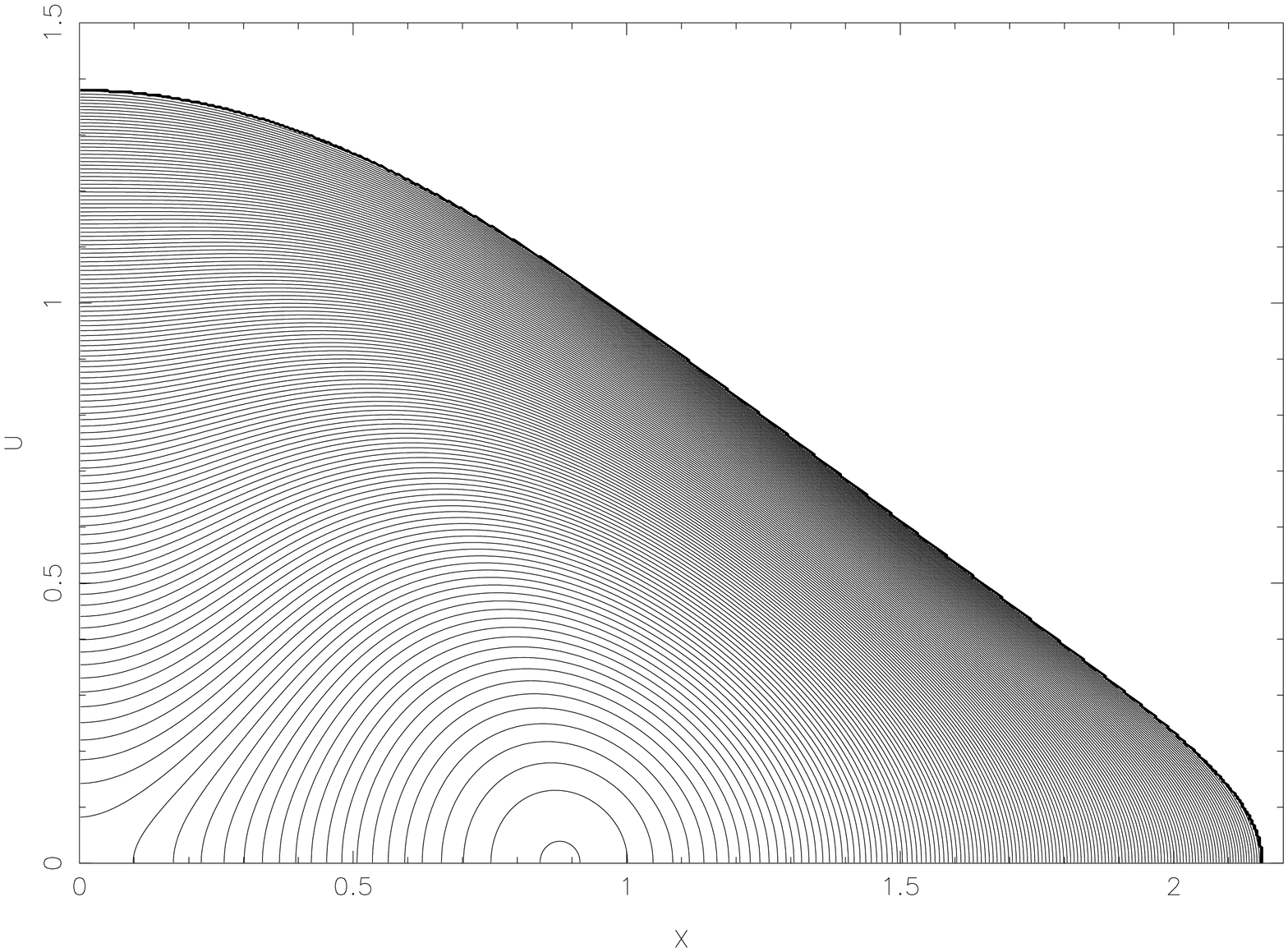}}}
\resizebox{8.5cm}{!}{\rotatebox{270}{\includegraphics{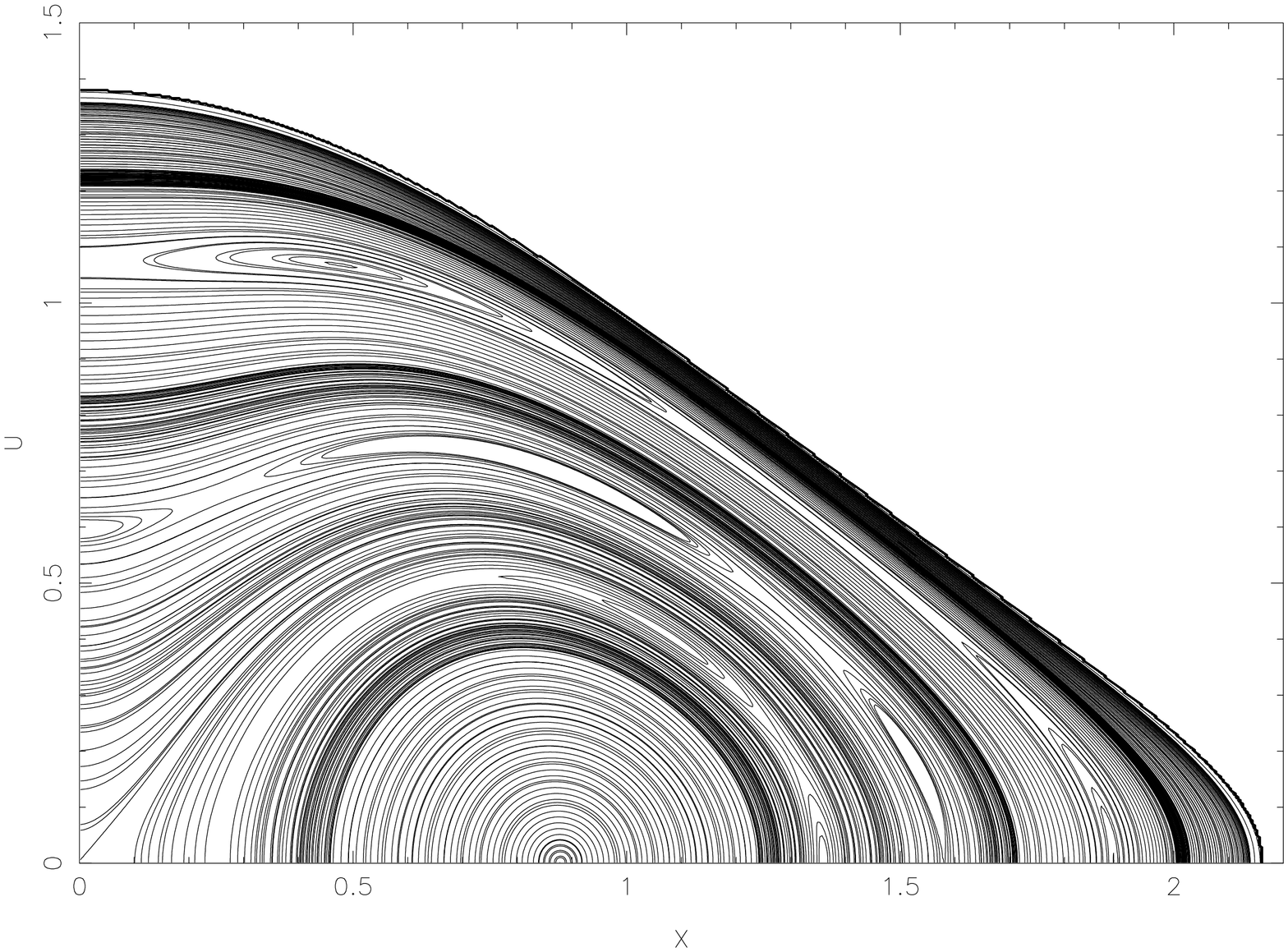}}}
\resizebox{8.5cm}{!}{\rotatebox{270}{\includegraphics{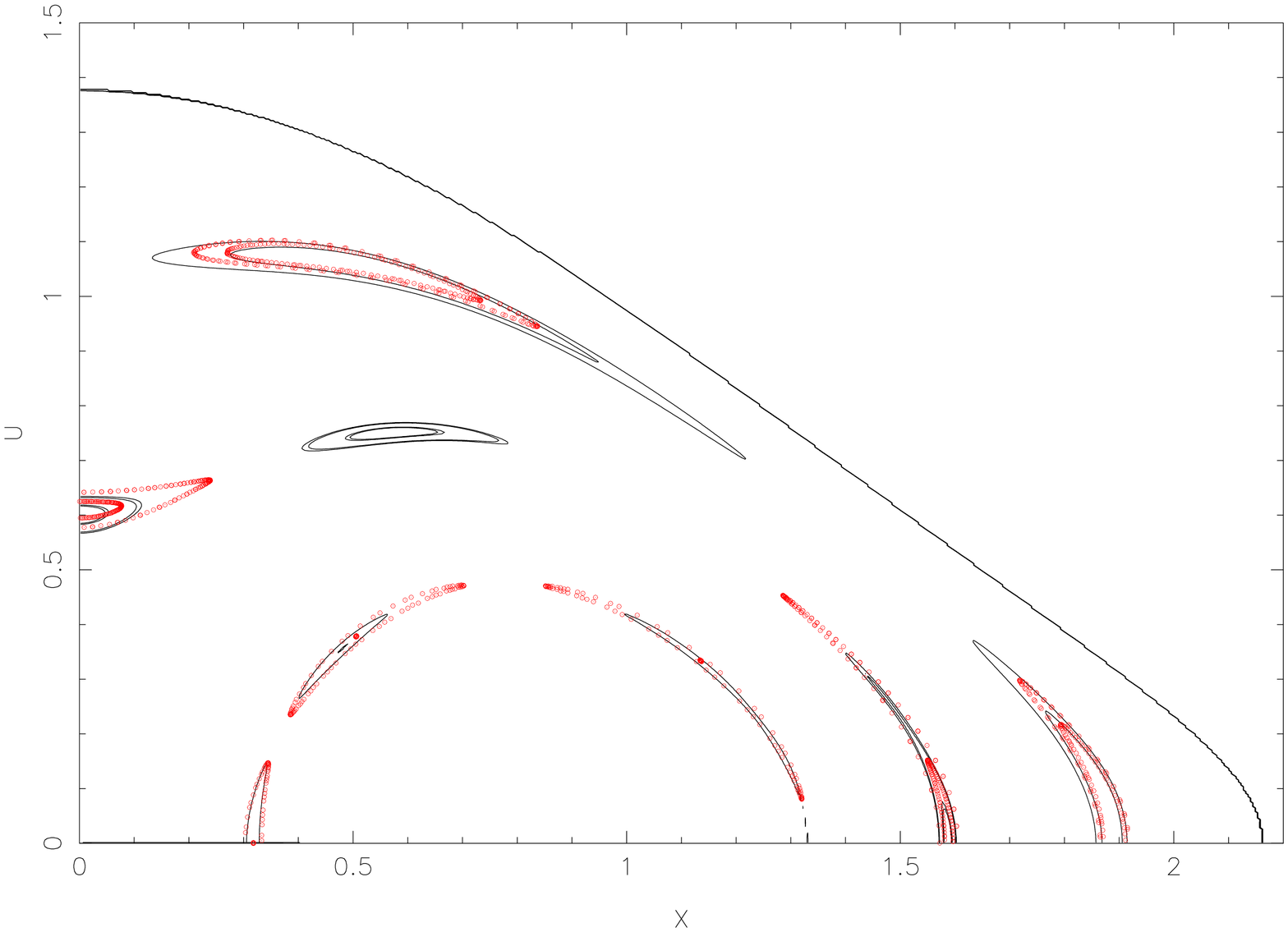}}}
\end{center}
  \caption{ - Exponential potential, $(x,u)$ Poincar\'e surfaces of section at $E=-0.05$ from the numerical and analytical $I$:
  (top-left) Numerical  section,
   (top-right) analytical section obtained from a fit at order 12 and  orbits at energies from $-1$ to $-0.05$,
   (bottom-left) from a fit at order 18 and orbits with energy $E=-0.05$,
  (bottom-right) from three local fits (order 18) of orbits in the neighbourhood of three resonant orbits, the empty space between the islands of stability is not filled correctly by the analytical integrals (red dots are the sections,  $y=0$ and $v>0$, from computed orbits). 
  }
    \label{fig5}
\end{figure*}

As a consequence of our specific choice of the cancelled coefficients within Eq.~1,
we have $I=0$  for  the radial orbits aligned along the  $y$-axis. On the other side,  the $x$-axis radial orbits are, at a given energy, the orbit with the largest $x$-apocentre, $X_{sup,E}$, and we have  $I$($X_{sup,E}$,0,0,0)=$X_{sup,E}^2$.  
  
  	In Figure~\ref{fig3}, the dispersion $\sigma_{I_{12}}$  is  also plotted. This dispersion  increases abruptly at the proximity of extended resonances. In Figure~\ref{fig4}, we plot $I_{12}(t)$ for 19 orbits close to a resonance 2:3  at $E=-0.05$. In this figure,  for each orbit $I_{12}(t)$ is plotted every $\Delta T=4$ from $T=0$ to 4000 (and each plotted orbit  is shifted for better visibility.) Crossing the resonance, the dispersion increases and  is of the order of the variation of $I$, and   within the resonance all  orbits have approximately the same mean value for $I_{12}$. As we  show below, this can be improved by increasing the number of fitted positions on orbits to improve the sampling of the various tori of orbits, and by increasing the order of the fitting polynomial to allow the modelling of tori with smaller scale structures.

Semi-ergodic orbits are also present and  are visible, confined between the tube and box orbits at $E=-0.05$ (Figure~\ref{fig5}).  At  energies $E=-0.05$, they correspond to orbits that pass close to the centre with $x_{\rm init}$ from 0 to $\sim$0.07.  Some of these orbits are included in the fit but do not  alter  the fitting quality: a  possible cause is that they remain  confined within a small volume and thus are reasonably adjusted  as the immediately nearby regular orbits. The dispersion of $I_{12}$ of the semi-ergodic orbits is significantly higher  than for the nearby regular  orbits, however.

Figure~\ref{fig5} (top-right) shows the Poincar\' e surface of section at $E=-0.05$ built from the fitted integral of motion $I_{12}$. We note that only two families of orbits are identifiable because no high-order resonances are  modelled.

\subsection{Higher orders}

To improve the approximate integral  using the same data within the same time interval $\Delta T=4000$, we progressively increase the order of the fitted polynomial  up to order 22 (2057 coefficients). By improving the fit, we  mean that we reduce  the variation of $I$ within a given time interval $\Delta T$. In contrast,  we  remark that   formal forms of the second integral obtained  also with polynomial series \citep{whi37, bir27, con60, gus66}  lead to polynomial series that are  asymptotic, i.e., up to an optimal order the accuracy of the series improves, and beyond this optimal order the accuracy decreases, the series being divergent.   This is related to the \cite{nek77}  theory \cite[see also][\S 2.3.6]{con02}, which establishes  that the formal integrals are valid over exponentially long (or short) times. 
Here, our approximate integral is different in the sense that it is always adjusted for a finite time interval, and its validity for longer time intervals has to be checked (the accuracy within the finite time interval can always be improved  up to a certain  limit $L$, and we expect that this limit $L$ will increase  when increasing the time interval, so that it does not  contradict with the Nekhoroshev theorem.)
With the 2D potentials considered here, the long-term behaviour of the orbits can be constrained by absolute barriers in the phase space due to KAM tori (at the opposite of the case of higher dimensional potentials that allow  Arnold diffusion). This may imply that our approximate integral variations would not increase for longer time.   
However, in the case of semi-ergodic orbits, orbits may be confined  within a restricted domain of the phase space before reaching another domain if the domains are scarcely connected, as illustrated by \citet[][figure 8]{ath83}. In such cases we would expect that the integral variations increase for  longer times.
\\

% figures 6

\begin{figure}
\resizebox{\hsize}{!}{\rotatebox{270}{\includegraphics{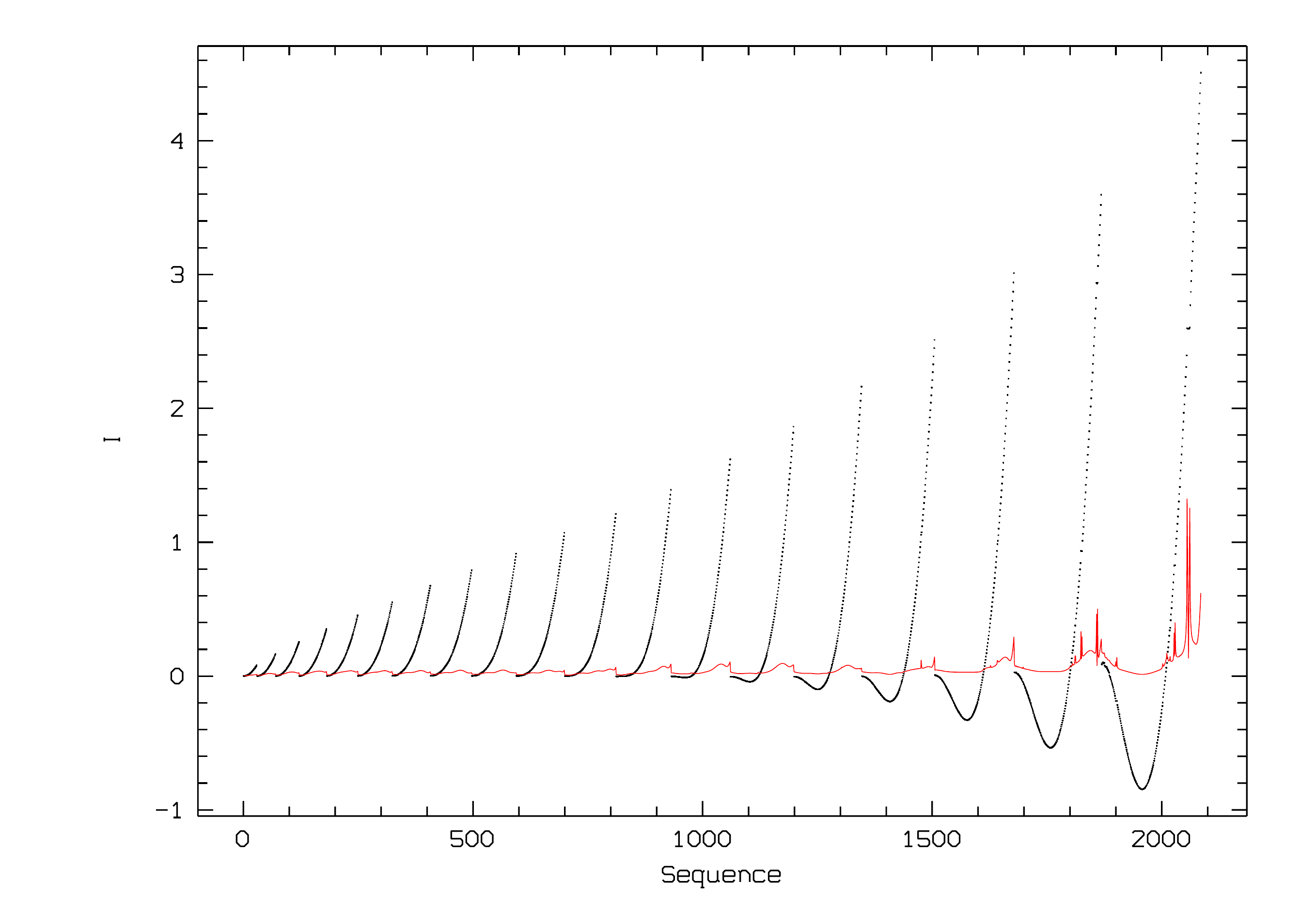}}}
\resizebox{\hsize}{!}{\rotatebox{270}{\includegraphics{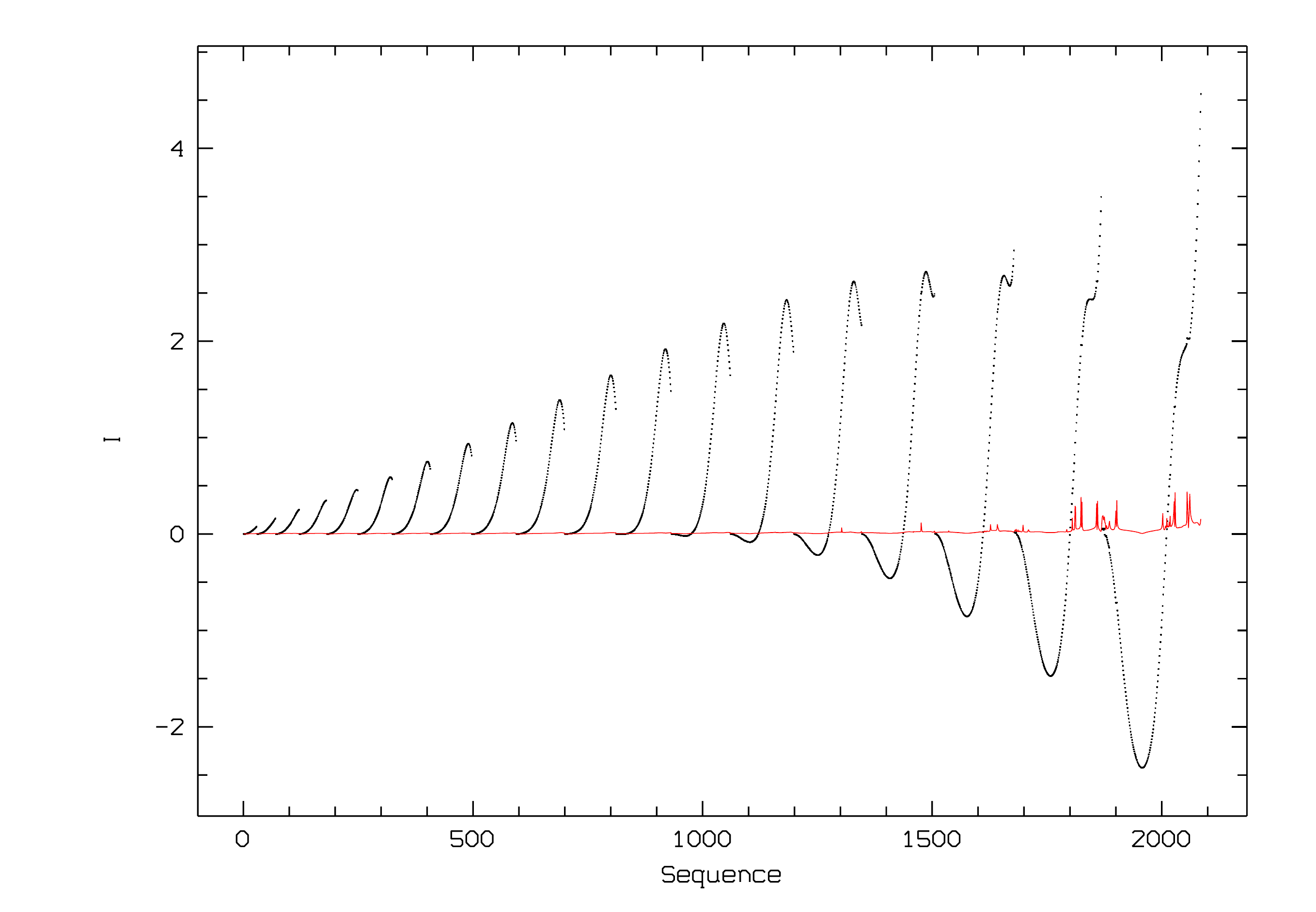}}}
\resizebox{\hsize}{!}{\rotatebox{270}{\includegraphics{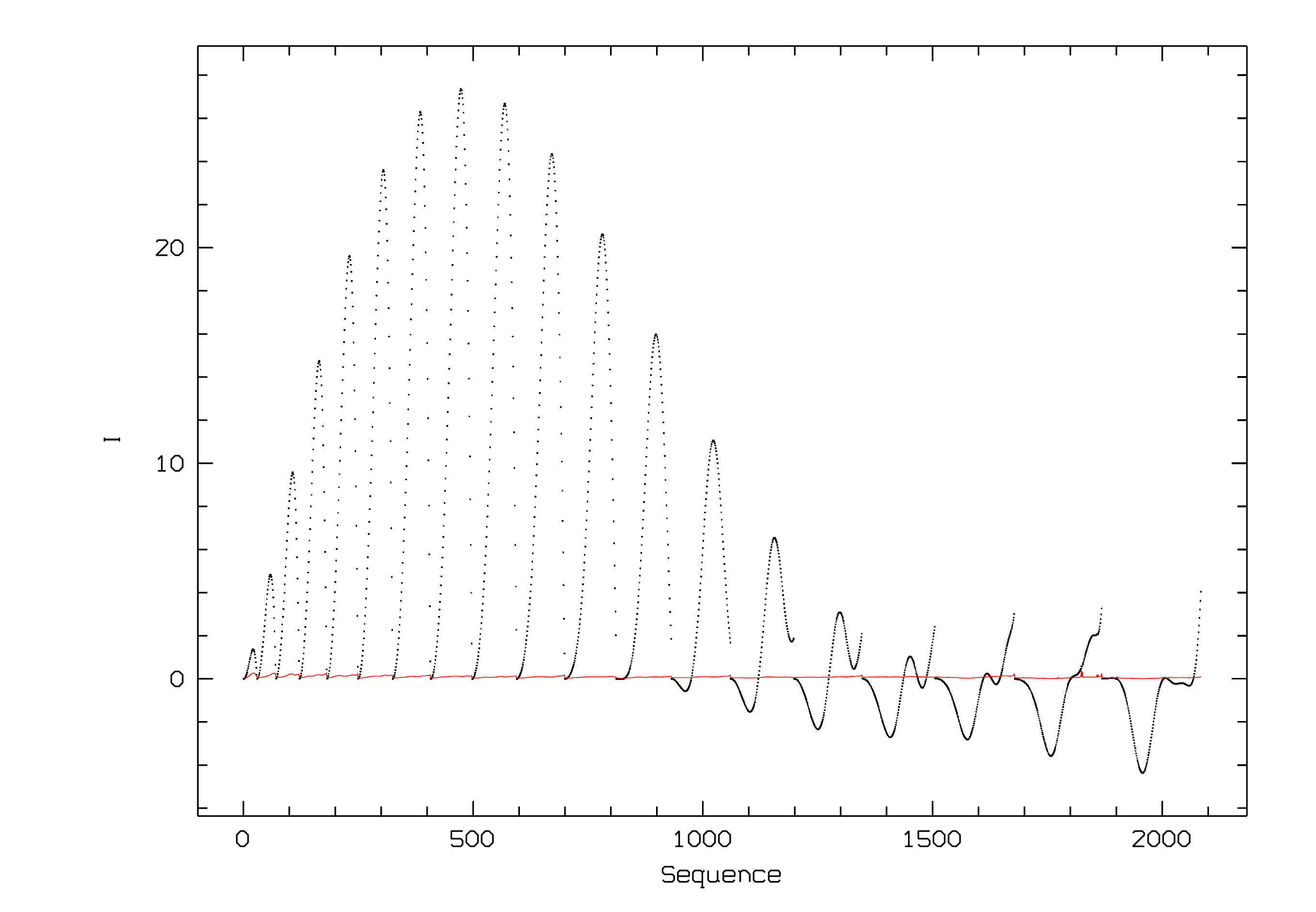}}}

\caption{ Integral $I$ for 2085 orbits covering 19 energies. From top to bottom, fits with polynomials of  order 10,16 and 22. Orbits  are plotted sequentially  according to their initial position $x_{init}$, and  according to the energy  ($E=-0.95$ left to $E=-0.05$ right). The red line is the dispersion of $I$ multiplied by 10.}
  \label{fig6}
\end{figure}

Figure~\ref{fig6}  shows the mean $I$ and dispersion $\sigma_I$ for each orbit, the values are presented sequentially by group of energy (left to right $E$ varies from $-0.95$ to $-0.05$). Increasing the polynomial order allows us to model  more resonances.  At order 10 the only one minimum of $I$ corresponds to the  resonant orbit 1:1.  At higher polynomial order, $I$ has more extrema   corresponding to  secondary resonances.  The dispersion through these resonances decreases progressively as the order of the polynomial increases: from  order 12 to 20  the mode (histogram maximum) of $\sigma_I$ changes from 0.050 to 0.0007, while the mean of $\sigma_I$  varies from 0.18 to 0.011. The mean dispersion increases at order 22, but the highest values of $\sigma_I$ at resonances  decrease significantly.

We define a more quantitative criterion to consider the ability to distinguish two nearby orbits with the same energy: the relative dispersion as  the ratio of the dispersion $\sigma_I$ to the variation of $I$ between  orbits: 

$\sigma_{rel}=  \sigma_I / |dI/dx|_E $. 

This relative dispersion has the  dimension of $x$, here with values of a few $10^{-4}$ and can be  compared to the full range of variations of $x$ from 0 to about 2. This relative accuracy is the poorest at resonances when $(dI/dx)_E$=0, but the mode of $\sigma_{rel}$ varies from $5.\, 10^{-4}$ to $1.5 \,10^{-4}$ when the polynomial order changes from order 12 to 20. The mean  $\sigma_{rel}$ (clipped within [0 to $1.\, 10^{-2}$]) decreases first but then flattens  at $1.0\, 10^{-3}$ from the orders 14 to 20. 

Increasing the polynomial order allows us to model more resonances. At low order from 8 to 16 only the two main families are modelled, as  is visible in the reconstructed Poincar\'e section at $E=-0.05$ (Figure~\ref{fig5}). At orders 18 to 22 three more resonances are visible in the reconstructed Poincar\'e surface of section corresponding to families 2:3, 3:4, and 3:5.  Only  the family  2:3  (extended from $x=0$, $u=1.06$ to $x=1.9$, $u=0$) is accurately modelled, the two other ones are  incorrectly positioned. Considering the small size of the structures of these resonances,  we suspect that a  higher order polynomial fit is necessary and also because  increasing the polynomial order   progressively improves the fit. 

Owing to  limited computing  resources we did not explore this systematically with much higher polynomial orders. Nevertheless, we explored fits with a restriction to orbits just close to the energy  $E=-0.05$  using  2200 orbits. The resulting fit is  better  (Figure~\ref{fig5}) with now two resonances correctly positioned within the surface of section, but only  the family 2:3 has  the correct multiplicity of islands. Now the family 3:4 is also correctly positioned, but the  number of reconstructed islands is incorrect, but by  increasing the polynomial order, the  exceeding island progressively disappears. 

Finally, we proceed to the local fit of the three resonances. Figure~\ref{fig5} shows the reconstructed Poincar\'e surface of section from the three analytical approximate integrals of motion that shows a satisfying fit. A second 1:4 family (not used for the fit) is also correctly modelled (the corresponding periodic orbit   never cross the $x$-axis perpendicularly).

All these results point out   that for this potential, an approximate integral is constant with a very  high accuracy.  There is also a strong indication  that a better modelling of $I$ could be achieved by using  a polynomial of sufficiently high order.\\

\subsection{Other combination of coefficients}

We have tried about 30 different combinations of coefficients to  numerically build quasi integrals $I$: all with the  same number of coefficients, and the fixed term is either $x^2$, $v^2$ or $x^2v^2$. The efficiency or accuracy of $I$ is determined according to the value $\sigma_{rel}$ for every orbit. The various combinations are not  equivalent in terms of efficiency,  some  give a better fit for orbits with high energies, others for low energies.
The four following combinations give the best results

- coefficient of $x^2$ set to 1, all others $v^{2n}$ and $x^{2k}$ are zero,  

- coefficient of $x^2$ set to 1, all others $y^{2l}$ and $x^{2k}$ are zero.

 With these  combinations we have  $$ I(0,0,0,0,v)=0 \,{\rm and}\, I(X_{sup,E},0,0,0)=X_{sup,E}^2$$

- coefficient of $v^2$ set to 1, all other $v^{2n}$ and $x^{2k}$ are zero,

- coefficient of $v^2$ set to 1, all other $v^{2n}$ and $u^{2m}$ are zero.

With these  combinations  
$$I(0,0,0,v)=2(E-\Phi(0,0))  \, {\rm and} \,  I(X_{sup,E},0,0,0)=0.$$

Applying the permutations of coefficients  $x \leftrightarrow y$ and also $u \leftrightarrow v$ gives  satisfying adjustments.

\section{Logarithmic potentials}

To check that the  numerical results described in the previous section  have some generality, we considered   other potentials. These   potentials  have  a very limited amount of ergodicity, the chaotic orbits remaining confined within a  small volume of the phase space. We also limited most of our fits using polynomials with order smaller than  18  to avoid prohibitively long computing times (100 h.cpu).

\begin{figure}
\resizebox{\hsize}{!}{\rotatebox{270}{\includegraphics{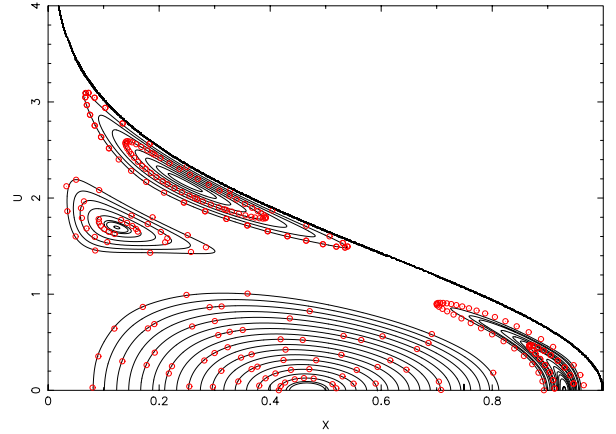}}}
  \caption{Logarithmic potential: ($x,u$)  Poincar\'e surface of section at $E=0$ for three families of orbits (circles) and three different analytical tori (continuous lines) obtained with polynomials of order 18.}
  \label{fig7}
\end{figure}

% 	FIGURE LOG

\begin{figure*}
\begin{center}
\resizebox{7.5cm}{!}{\rotatebox{270}{\includegraphics{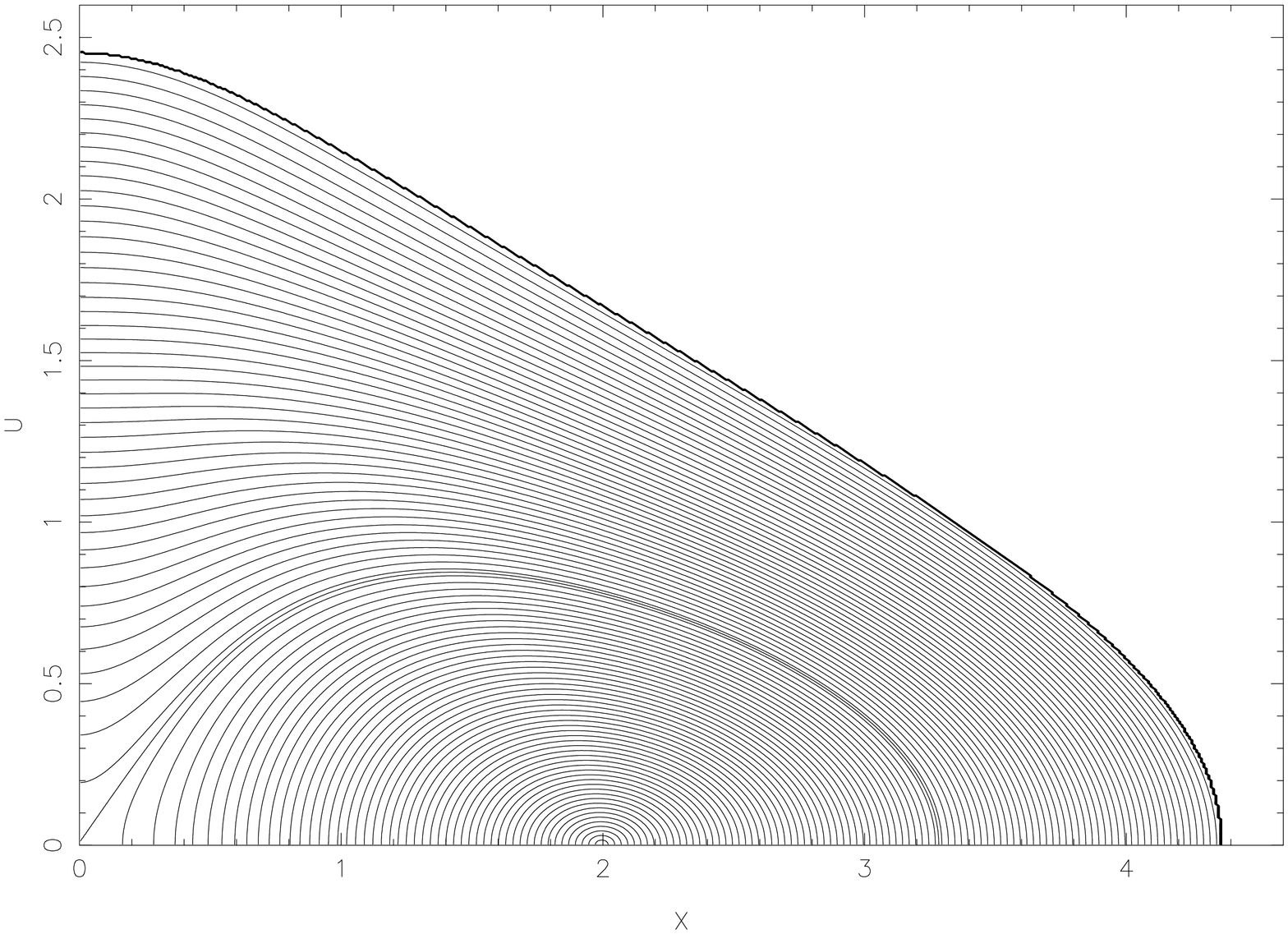}}}
\resizebox{7.5cm}{!}{\rotatebox{270}{\includegraphics{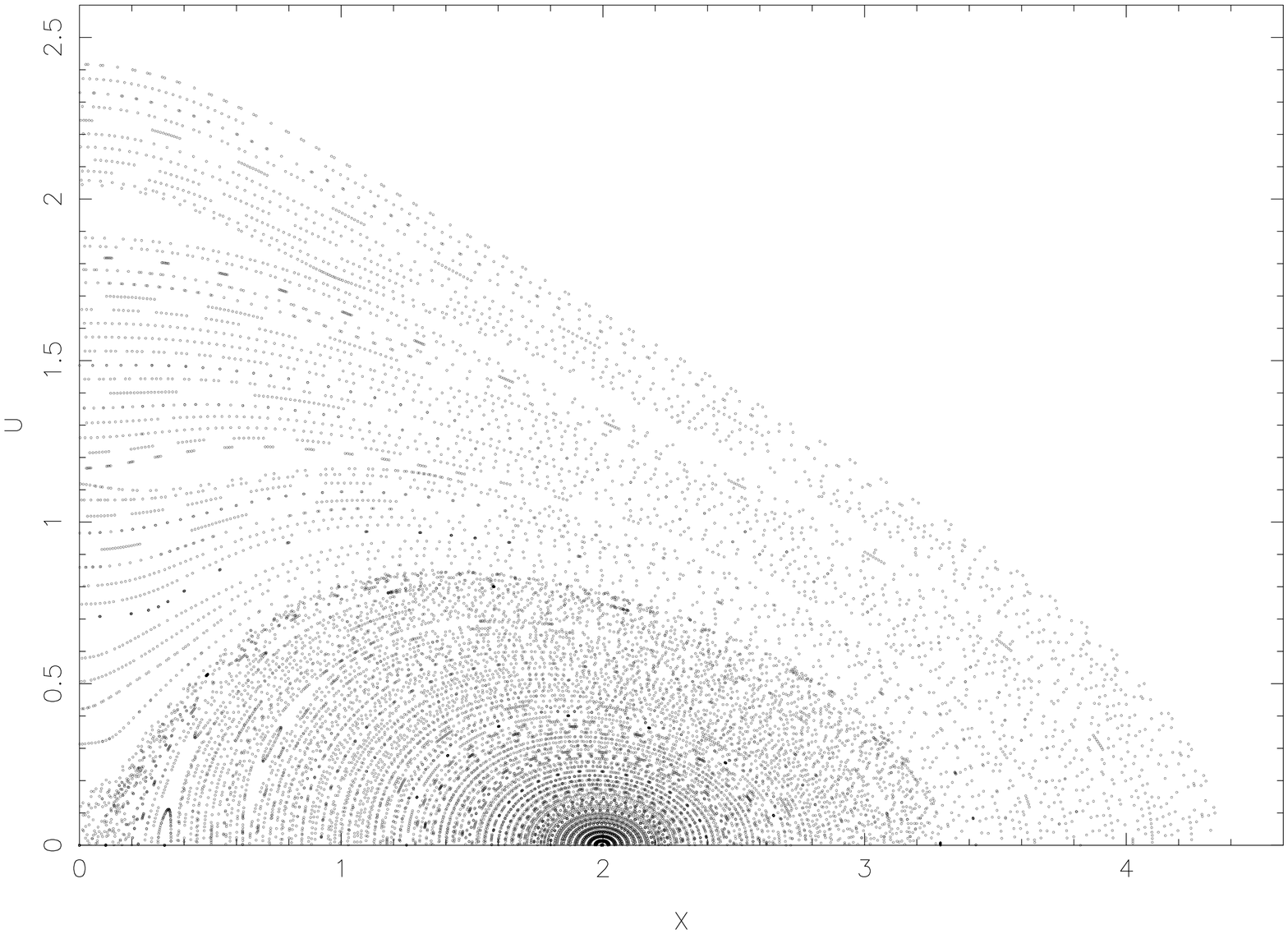}}}
\resizebox{7.5cm}{!}{\rotatebox{270}{\includegraphics{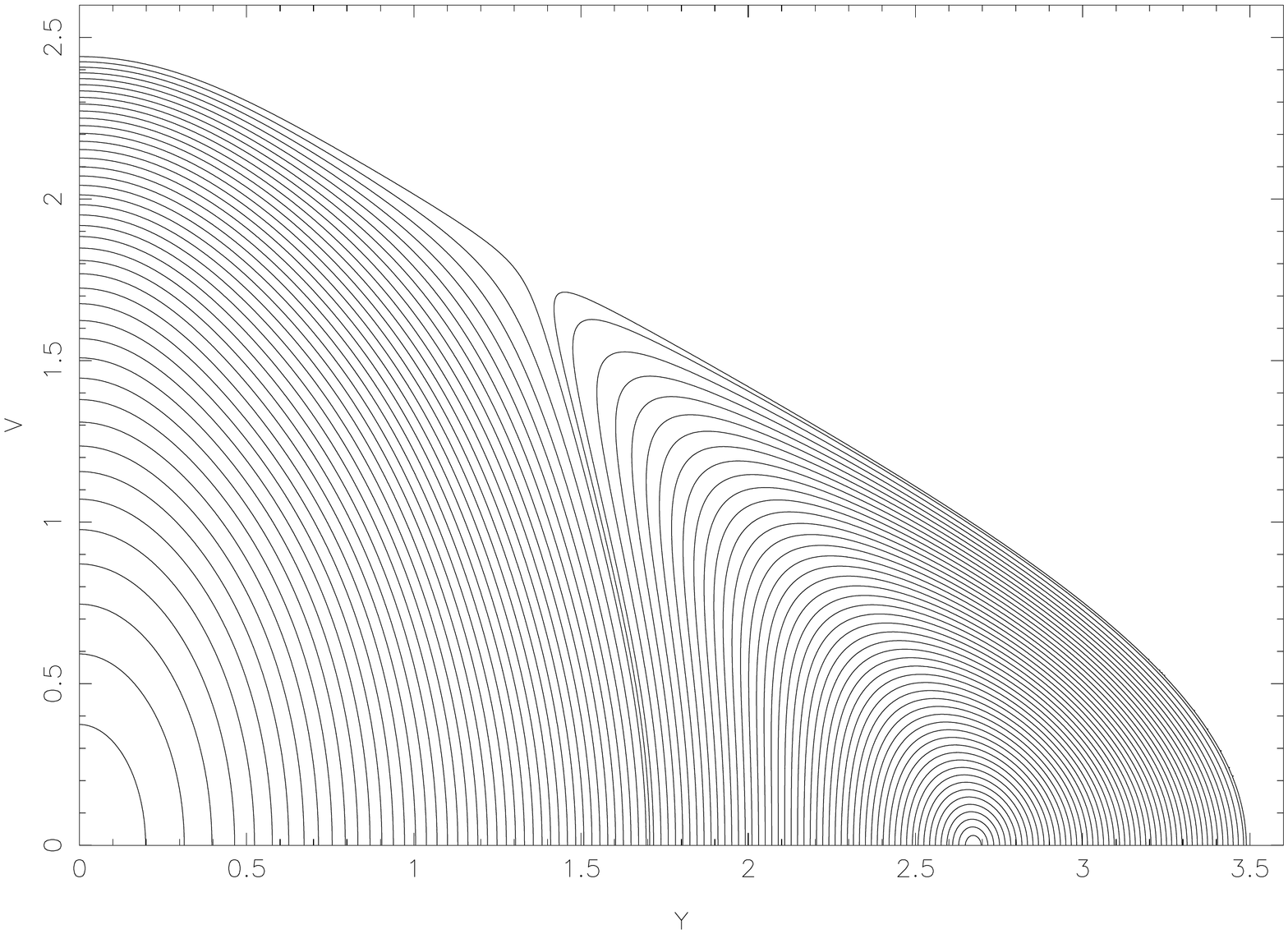}}}
\resizebox{7.5cm}{!}{\rotatebox{270}{\includegraphics{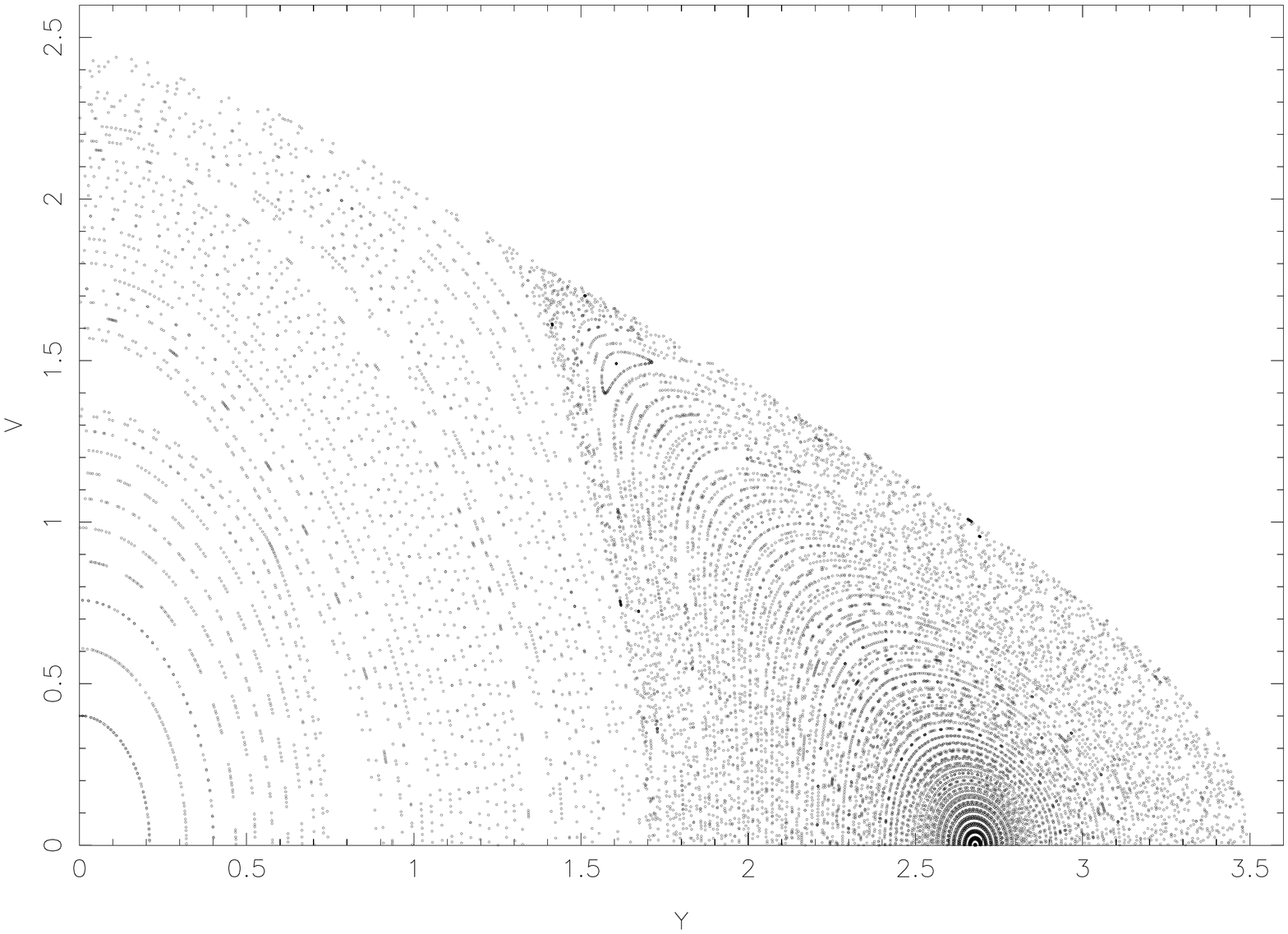}}}
\end{center}
  \caption{Logarithmic potential with a core: ($x,u$) (top) and  ($y,v$)  (bottom) Poincar\'e surfaces of section at  $E=3$. Fits with an $18^{th}$ order polynomial from orbits with $E=0$ to 3.5. Computed orbits (right) and anaytical tori (left).  }
  \label{fig8}
\end{figure*}

\subsection{Scale-free logarithmic potential} 

The logarithmic potential introduced by \cite{ric80}  to model axisymmetric disc galaxies  produces a flat rotation curve:

$$ \Phi= \log( x^2 + y^2/q^2) . $$

 The parameter $q$ allows us to model a flattening of the potential and of the corresponding density. The density is not  positive everywhere when  with $q< \sqrt{2}/2 $, and  that with $q \le \sqrt{3/4} \sim 0.87$   the density isocontours present a depression close to the $y$-axis.

A detailed analysis of orbits in this potential was presented by \cite{mir89} with plotted Poincar\'e surfaces of section.  This potential has semi-ergodic motions \citep{pap96} and is   not integrable.  Chaotic orbits   have excursions close  to the centre and  circulate around   the sets of tori of the major families of resonant orbits. The phase space is mainly occupied by regular orbits (at least with $q$ from 0.2 to 1) and the dominant families of these regular orbits are  tube orbits of type 1:1.  Box orbits of type 2:1, 3:2, 4:3, etc (also named boxlets orbits : banana, fish, pretzel, etc)  cover most of the remaining surface of the Poincar\'e section.

Using the procedure described in  Section 2, we determined polynomial forms to evaluate a second integral of motion by fitting   orbits  of type 1:1, 2:1, and 3:2 (this is done only  with $E=0$ since the potential is scale-free and   the integral can be simply deduced for other energies). Using polynomials of order 18, we succeeded in  modelling  most of the orbits of each of these three families separately. Thus, a different polynomial was used for each of the three families (Figure~\ref{fig7}).  We also obtained a general fit of the three families with a unique polynomial of order 28 (4760 coefficients), but just two-thirds of the orbits plotted in Figure  \ref{fig7} are fitted (the orbits close the central periodic orbits). Within the surface of section ($x$, $u$),  these unfitted orbits  are  poorly fitted when $x \la 0.1$ but are correctly fitted at larger $x$.

 The scale-free logarithmic potential is the one for which the application of our method gives  the most limited success. As for the other potentials, we used our program to fit the energy $E(x,y,u,v)$  considering it as a polynomial and consequently also the potential $\Phi(x,y)$. Not surprisingly, a polynomial form hardly fits a logarithmic function without core. Furthermore,  examining the Boltzmann equation for this potential, we can assume that to build  a second integral with a formal series, a series of rational functions will be  more suited. Here,  the \cite{pre82} method to obtain rational solutions should be certainly more efficient.
  
\subsection{Logarithmic potential with a core}

As suggested by \cite{ric80} and already used by \cite{bin81}  and \cite{mag82}, the logarithmic potential 

$$ \Phi=\log (a^2+x^2 + y^2/q^2)- 2\log(a)$$

has a core. We examine this potential with   $a=1$ and $q=0.8$. The presence of a core  drastically reduces the extent of the semi-ergodic orbits for energies $E<4$ (and $a=1$). It is  at much  higher energies and  with $x$ or $y$ coordinates much larger than the core radius $a$, that  orbits  are  similar to the orbits within the logarithmic potential seen in Section 4.1. At low energies box orbits dominate, type 1:1 tube orbits appear at $E\sim$ 0.9.

We obtain a  good fit for orbits from $E=0$ to 3 with a $12^{th}$ order polynomial (and coefficients selected as in Section 3.1), the fit is good everywhere except close to the transition between box and tube orbits, the less accurately fitted region being  at the apparition of tube orbits when $E\sim$ 0.9.  The fit is greatly improved, however, by increasing the number of orbits in this transition region, for instance  using 23428 orbits (1000 positions along each orbit) and order 16. Some resonant orbits are not modelled (see Figures~\ref{fig8}, $(y,u)$ Poincar\'e section  at $E=3.0$, and $(y,v)$=(0.8,0) or (0,1.5) or (1.7,1.5)) but they cover a very small volume of the phase space. Increasing the order to the $18^{th}$ order, the fit improves,  but  high-order resonances are still poorly fitted.
Fitting orbits from $E=0$ to 3.5 (with 27619 orbits) with 5000 positions  on each orbit and order 18, the fit improves from $E=0$ to 3.

\subsection{Axisymmetric  scale-free logarithmic potential}

\begin{figure}
%\resizebox{8.6cm}{!}{\rotatebox{270}{\includegraphics{fig9a.pdf}}}
%\resizebox{8.6cm}{!}{\rotatebox{270}{\includegraphics{fig9b.pdf}}}
\resizebox{\hsize}{!}{\rotatebox{270}{\includegraphics{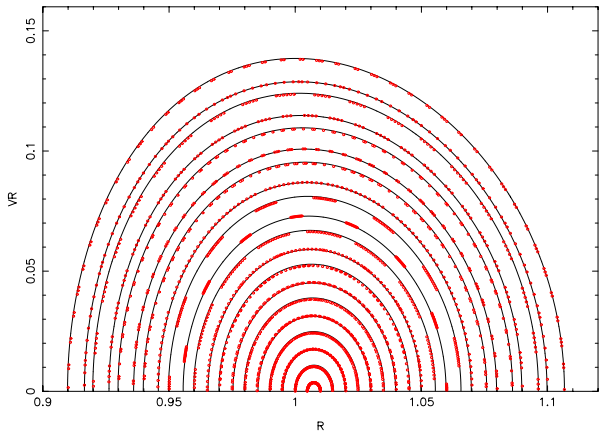}}}
\resizebox{\hsize}{!}{\rotatebox{270}{\includegraphics{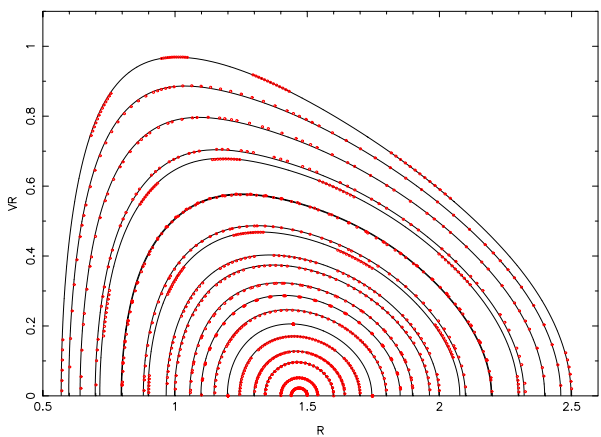}}}
  \caption{Axisymmetric logarithmic potential: orbits with $L_{\rm z}=1$. (top to bottom): ($x,u$)  Poincar\'e surface of section with  $E=0.01$ and .5.  Computed orbits (dotted red circles) and analytical reconstructed tori (continuous dark lines) with an $18^{th}$ order polynomial fitted along orbits with $E=0$ to 0.5.}
  \label{fig9}
\end{figure}

Still following \cite{ric80}, we consider a 3D flattened  logarithmic potential and motions with non-null angular momentum. We set no core ($a=0$) and  a flattening parameter $q=0.8$  and  consider only  orbits with the same angular momentum $L_z=1$. Results for any other non-null  $L_z$  are easily deduced.
Thus we consider orbits in the effective potential

$$\Phi_{\rm eff}= \frac{v_c^2}{2} 
\log( R^2 + z^2/q^2) +\frac{L_z^2}{2\,R^2}  -   \frac{v_c^2}{2}(\log L_z + \frac{1}{2}).$$

$E_{\rm eff}$ close to zero corresponds to nearly circular obits around the $z$-axis. The radial orbits with $z=0$ and a null vertical velocity dispersion $\sigma_{z}$ are a first family of periodic orbits. The shell orbits with a null radial velocity  ${v_R}$ when $z=0$ are another family. For a given $E_{\rm eff }$, the amount of radial versus vertical extension of an orbit is constrained by a second integral \citep[see for instance][]{oll62}. The rotation prevents orbits from circulating close to the centre and thus reduces the amount of ergodic orbits by comparison to the case with  $L_z=0$. Chaotic orbits  occupy  an extremely small volume in the phase space, at least for $E_{\rm eff} < 8$. At low energies ($E \la 1$) orbits are mainly box orbits. 

 Radial orbits have $z=0$ and  at their  two  extrema $R_{\rm min}$ and $R_{\rm max}$  have also $v_R=v_z=0$. Thus, for all such orbit: $I(R_{\rm min},0,0,0)=I(R_{\rm min},0,0,0) $, a relation that  can be achieved only if $I(R,0,0,0)=f(\Phi(R,z=0))$. The choice of coefficients in Section 3.1 does not allow us to fulfil this condition. Therefore here  we set to one the coefficient of $v_z^2$ and remove all $v_z^{2n}$, with $n > 1$, and all $v_R^{2m}$ terms (removing all $v_z^{2n}$, with $m > 1$, and all $R^{2k}$ is also satisfying.)
 
Figure~\ref{fig9} shows the results of a fit with 24000 orbits (100 positions on each orbit) with $E=0$ to 0.5, and an $18^{th}$ order polynomial.

% FIGURE TODA LATTICE

\begin{figure}
\resizebox{\hsize}{!}{\rotatebox{270}{\includegraphics{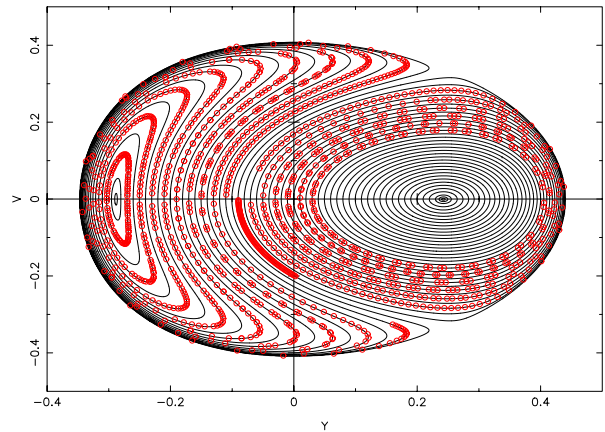}}}
\resizebox{\hsize}{!}{\rotatebox{270}{\includegraphics{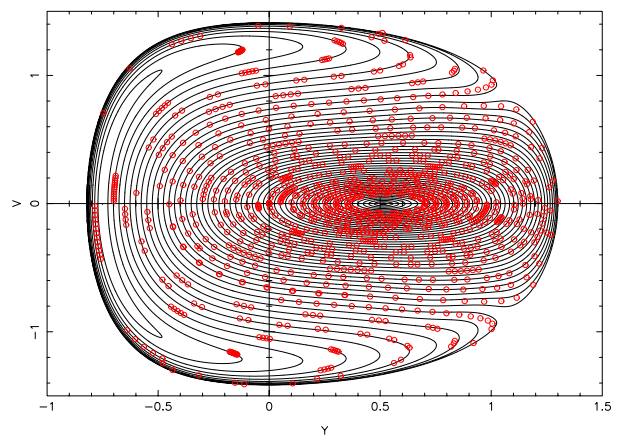}}}
  \caption{Toda lattice: (Top) $(y,v)$  Poincar\'e section at ($x$=0, $u>0$) and $E$=1/12 and analytical tori (continuous  lines) with a 14$^{th}$ order polynomial.  (Bottom)  ($y,v$)  Poincar\'e section at ($x$=0, $u$$>$0) and $E=1$ for a few orbits (dotted red circles) and analytical tori (continuous dark lines) with a 16$^{th}$ order polynomial.
   }
  \label{fig10}
\end{figure}

% FIGURE HENON & HEYLES

\begin{figure}
\resizebox{\hsize}{!}{\rotatebox{270}{\includegraphics{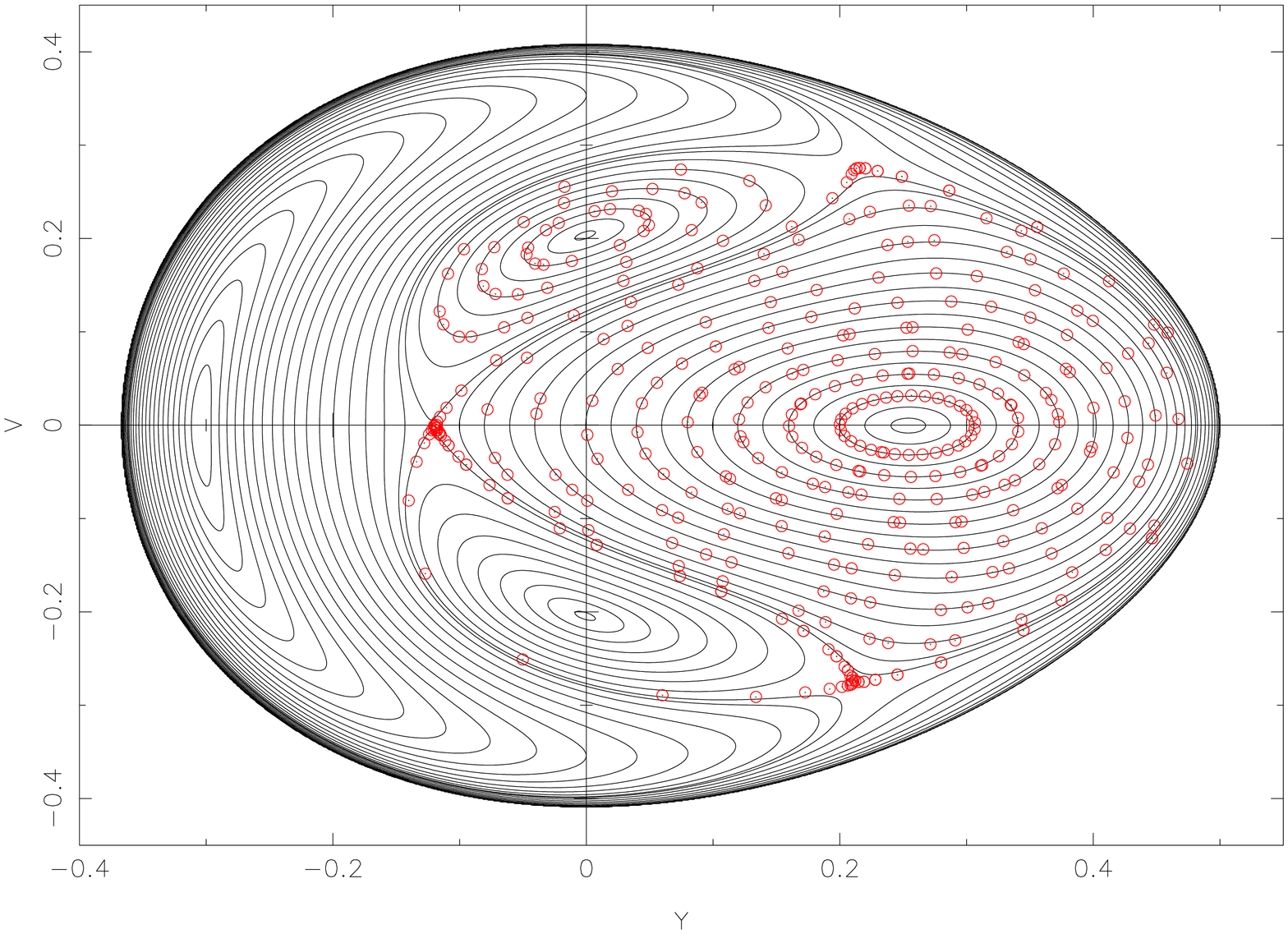}}}
\resizebox{\hsize}{!}{\rotatebox{270}{\includegraphics{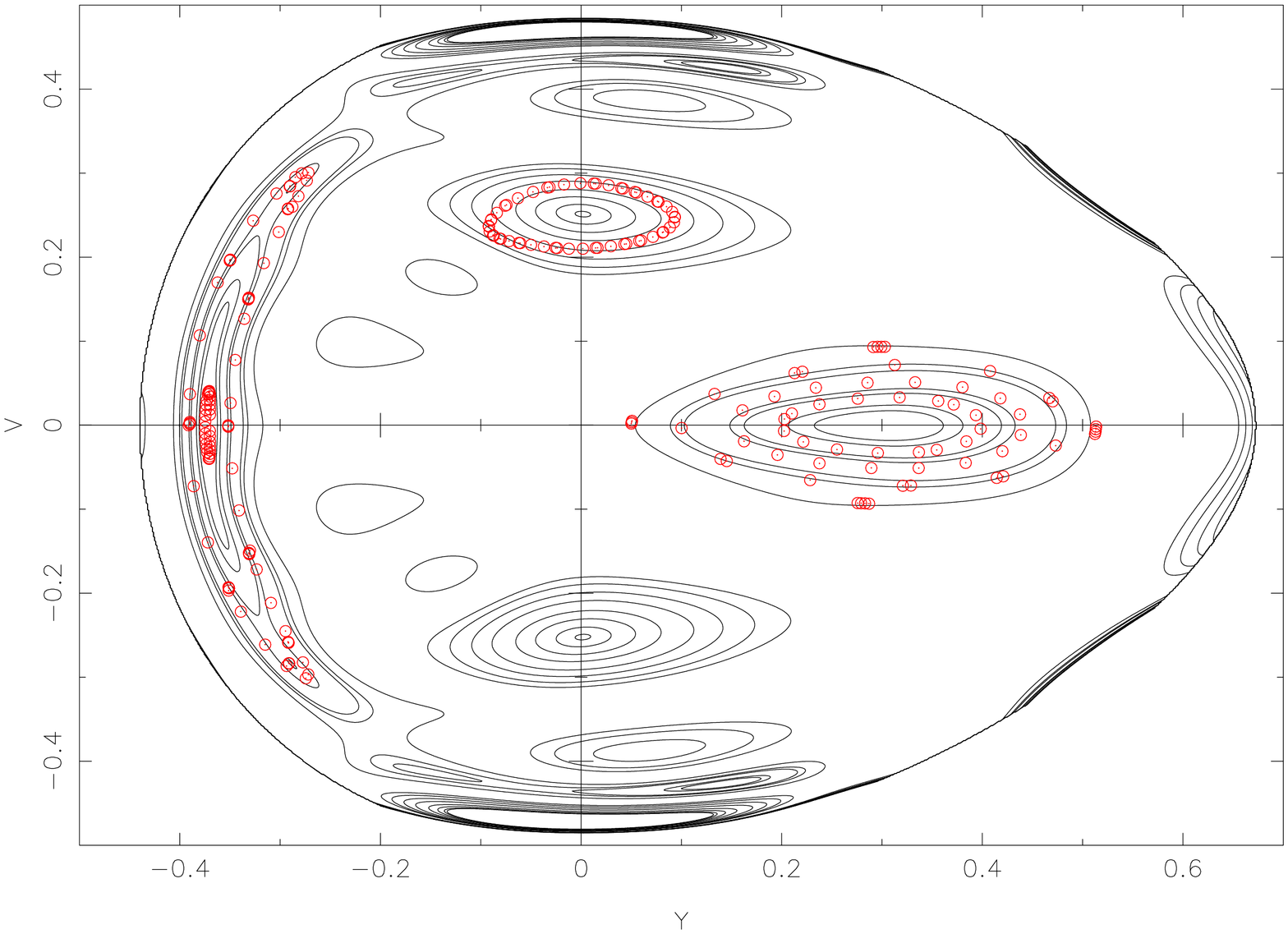}}}
  \caption{ H\'enon \& Heiles potential: (Top) ($y,v$)  Poincar\'e section at ($x$=0 and $u>0$) and $E=1/12$ for a few orbits (dotted red circles) and analytical tori (continuous dark lines) with a 12$^{th}$  order polynomial.
  (Bottom) Poincar\'e section at  $E=1/8$ for a few regular orbits (dotted red circles) and analytical tori (continuous dark lines) with an $18^{th}$ order polynomial, the unplotted domain is covered by chaotic orbits.
  }
  \label{fig11}
\end{figure}

\section{H\'enon \& Heiles and  Toda lattice potentials}

We do not present the various tests of our procedure performed  with  integrable dynamical systems:  axisymmetric,  St\"ackel potentials, or systems with a second integral with a finite polynomial form.

Instead, we  summarise  the results  obtained for two classic potentials, the \cite{tod70} lattice (2D case) and the \cite{hen64}  potential \cite[a historical description concerning these potentials on simulations in dynamics is given by][]{wei97}. The Toda lattice potential  has a known analytical second integral \citep{hen74, man74, man75,fla74}. The   H\'enon \& Heiles potential  has regular orbits at low energies while nearly full ergodicity appears at   high energies. The two potentials are  closely linked since the first four terms of the series development of the 2D Toda lattice   give  the H\'enon \& Heiles potential, one is integrable, the other one shows  a nearly complete dynamical chaos at  high energies. The functional form of the second integral of the Toda lattice is known and is odd for the momenta reflecting that there is no box orbit but only circulating motions (tube orbits). The extrema of the second integral at fixed energy allows the precise identification of the     family of stable orbits.  This potential and  its  second integral can  be written as  a polynomial series with an infinite radius of convergence for all coordinates. This was assumed to be important as a first step to test the procedure developed in this paper, which  assumes that a polynomial series exists to represent the hypothetic second integral.\\

\subsection{Toda lattice}
Selecting 79 orbits with  energy $E=1/12$, 10000 positions on each orbit, and a fit at order 14, we obtain a median relative accuracy of 1.2\,10$^{-5}$. We obtain  7.5\,10$^{-7}$ at order 16 (Fig.\,\ref{fig10}).  Selecting a uniform distribution of 1700 orbits from energy $E=0$ to $1$, still with 10000 positions, we obtain a median relative accuracy of 1.5\,10$^{-1}$ at order 14 and 3.5\,10$^{-2}$ at order 18. We also succeed in obtaining  a good fit from $E=0$ to 4 with 2800 orbits. The polynomial expansion of the known Todda lattice second integral contains only terms of order 0, 1 or 3 in  the momenta $v$. Our {\it a priori} construction and fit of a polynomial series does not impose such a constraint  on the coefficients, accordingly our fitted series requires a much larger number of terms and coefficients to achieve the same accuracy as using the  polynomial form of the known integral. 

\subsection{H\'enon \& Heiles potential}

Within the H\'enon \& Heiles potential, we  compute   10000 positions along 41 orbits with the energy $E=1/12$  ($\Delta$T=10000). The recovered analytical tori are quantitatively correct and better than a few thousandths in relative accuracy  at orders higher than the $7^{th}$ (83 coefficients). Figure\,\ref{fig11} may be compared to the result of \citet[][ figure 13]{gus66}, where the integral of motion was determined  by symbolically manipulating polynomials  up  to the order 7. At order 12, our median relative accuracy  is improved:  $2\,10^{-4}$, and at order 14 it is  $10^{-4}$.
 
At energy $E=1/8$ an ergodic orbit covers about half of the phase space 
\cite[see figure\,5 in][]{hen64}. We obtain  a roughly  approximate integral of motion for the regular orbits close to the main periodic orbits. With  a fit of 54 regular  orbits close to the  resonance 1:1,   the recovered  analytical tori (Fig.\,\ref{fig11})  show that the three main families are recovered. This  may be compared to the results of \citet[][figure 10]{gus66} obtained at the same energy: he obtained analytical sections in good qualitative agreement with the sections plotted from computed orbits.  More recent works \cite[see][]{kal92,rob93,con03} give higher orders of formal integrals and show a  better agreement at  energy E=1/8 where the chaotic orbits occupy a large volume of the phase space.

\section{Conclusion}

Obtaining integrals of motion of a dynamical system   simplifies the analysis and the understanding of the system,  because  it allows us for instance to build more easily tractable distribution functions. It  also gives a synthetic description of the orbits, the building blocks of galaxies. This is well-known for axisymmetric systems, and it has been also extensively developed  for St\"{a}ckel potentials. Other  systems with known analytic integrals are apparently  less interesting for the astronomical community.

Therefore many efforts and methods have been developed to obtain  approximate integrals of motions for more general potentials. Moser's theorem states that most invariant curves will be preserved under a weak perturbation of an integrable dynamical system.
A method initiated by \cite{bir27}  consists of 
writing the Hamiltonian in the so-called  normal form, which  allows the formal construction of polynomial series  that are solutions of the Boltzmann equation.These series are generally divergent \citep{sie41, con03}.  Even for an integrable system, they may be   not  convergent \citep{woo87}. \cite{gus66} developed an algorithm to obtain these formal integrals and applied it to the H\'enon \& Heyles potential.

	A more direct  method by \cite{whi37}  or \cite{con60} also consisted of  looking for formal polynomial integrals of motion, but did this directly by substituting and comparing coefficients.   \cite{gio78} solved the consistency problem of these direct methods.

Inspired by these methods,  \cite{gio78} proposed a new method for constructing formal integrals near an equilibrium point, and  a  numerical program  \citep{gio79}  was used for applications  \cite[see for instance][]{kal92, con03}. The formal integrals are  different  for resonant and each non-resonant orbit \citep{con00}. 

 A  different  technique, called torus construction \citep[for a review, see][]{val99}, consists of   numerically  mapping  the action-angle coordinates of a known potential into the action-angle coordinates for the system under investigation \citep{mcg90}. This is obtained  by a least-squares procedure. It implies the availability of a close and integrable Hamiltonian whose tori can be simply mapped to the target tori; it also implies mappings from different toy tori for different orbit families \citep{kaa94a}. Moreover, there are general methods of semi-numerical perturbation that take into account the full 'distortion' of the invariant tori \citep{hen90}.

The  method developed in this paper partly builds on  previously published methods.  We solved the Boltzmann equation with a polynomial, but numerically  and with a fit to many  peculiar solutions (i.e. orbits). This method is {\it a priori } limited by the accuracy with which the invariant curves are defined on the surfaces of section, i.e. to which point they are effectively invariant and affected by diffusion. This limiting accuracy depends, for a given potential and a given orbit, on the time interval of integration of the orbit. But for a given time interval, it should be sufficient to increase the order of the fitting polynomial to achieve higher accuracy.  We succeeded  in different examples  to fit  various resonant and non-resonant families of orbits  with the same integral. The quality of the fits depends  on selected coefficients of a polynomial series.
Here, the method  was applied to various 2D potentials  representative of motions within elliptical galaxies or  motions within a non-axisymmetric disc.  For all these potentials, a unique analytical integral of motion was obtained for an extended range of energies and  regular orbits. The coefficients of the polynomial  of the integrals were obtained from a linear least-squares minimisation.
\cite{war91} and \cite{baz91} also
developed   methods of orbit fitting but with   limitations; the former with restrictions   to high-order resonant orbits or non-resonant orbits, the latter to 2D sections. In addition, they did not impose the energy independence of their integral, which may  impact the numerical accuracy of the fitted second integral.

As a future development, the method presented in this paper will be applied  to the construction of a  simple 3D galactic model. We will use a scale-free logarithmic disc-halo potential \citep{bie09}, for which a third integral $I_3 $ will be determined:  the third integral is first determined for a given value of the energy and a given non-zero angular momentum,  and   is then simply deduced for other values of $E$ and $L_z$. Afterwards,  the  action  can be evaluated from $I_3$. Finally, a distribution function for an exponential stellar disc  can be built following the generalisation of the Shu distribution function  by \cite{kui95}, \cite{bie97},  or \cite{bie99}. This model will allow us  to  explore  various properties: for instance, to check the domain of validity of the asymmetric drift relation especially out of the mid-plane, to examine the velocity ellipsoid tilt  dependence on the distribution function, or to build a realistic model to measure the $K_z$ force from data far above and below the galactic plane by minimising  the number of free parameters and assumptions in the modelling.

% for the bibliography, at the end 
\bibliographystyle{aa} % style aa.bst 
\bibliography{VersionNov3} % your references Yourfile.bib

\end{document}